\documentclass[conference,10pt]{IEEEtran}

\usepackage[utf8]{inputenc}
\usepackage{booktabs}
\usepackage{graphicx}
\usepackage{subfigure}
\usepackage{units}
\usepackage{blindtext}
\usepackage{xcolor,xspace}
\usepackage{amsmath}
\usepackage{amssymb}
\usepackage{paralist}
\usepackage{amsmath, amssymb}
\usepackage{tikz}
\usepackage{booktabs}
\usepackage{pifont}
\usepackage{multirow}
\usepackage{hhline}
\usepackage{balance}
\usepackage{footnote}
\usepackage{logicproof}
\usepackage{amsthm}
\usetikzlibrary{patterns}
\usetikzlibrary{calc}
\usetikzlibrary{decorations.pathreplacing}
\usetikzlibrary{arrows,automata}
\usetikzlibrary{arrows,calc,shapes,decorations.pathreplacing}
\usepackage{pbox}
\usepackage{tabularx}

\usepackage{arydshln}
\usepackage{mdframed}
\usepackage{pgf-umlsd}

\usepackage[T1]{fontenc}
\usepackage{beramono,textcomp}

\newcommand{\new}[1]{\textcolor{black}{#1}}

\usepackage{listings}
\usepackage[ruled]{algorithm2e}
\usepackage{algorithmic}

\LinesNumbered

\usepackage{url}

\expandafter\def\csname ver@l3regex.sty\endcsname{}
\usepackage[n,advantage,operators,sets,adversary,landau,probability,notions,logic,ff,mm,primitives,events,complexity,asymptotics,keys]{cryptocode}

\newcommand{\ppstext}{{Verified Remote Sensing Authorization}\xspace}
\newcommand{\ppstextunderlined}{{\underline{Ve}rified \underline{R}emote \underline{S}ensing \underline{A}uthorization}\xspace}
\newcommand{\pps}{{\texttt{VERSA}}\xspace}
\newcommand{\pfbtext}{{\texttt{Privacy-from-Birth}}\xspace}
\newcommand{\pfb}{{\texttt{PfB}}\xspace}
\newcommand{\atoken}{{\small \texttt{ATok}}\xspace}

\newcommand{\operation}{{state \xspace}}
\newcommand{\operations}{{states \xspace}}

\newcommand{\vrased}{{{\sf\it VRASED}}\xspace}
\newcommand{\attkey}{\ensuremath{\mathcal K}\xspace}
\newcommand{\enckey}{\ensuremath{\mathcal K_{enc}}\xspace}

\newcommand{\pass}{\ensuremath{\inst_{Auth}}\xspace}

\newcommand{\ermin}{\ensuremath{\ER_{min}}\xspace}
\newcommand{\ermax}{\ensuremath{\ER_{max}}\xspace}

\newcommand{\daddr}{\ensuremath{D_{addr}}\xspace}
\newcommand{\ren}{\ensuremath{R_{en}}\xspace}
\newcommand{\wen}{\ensuremath{W_{en}}\xspace}
\newcommand{\dmaaddr}{\ensuremath{DMA_{addr}}\xspace}
\newcommand{\dmaen}{\ensuremath{DMA_{en}}\xspace}
\newcommand{\irq}{\ensuremath{irq}\xspace}
\newcommand{\pc}{\ensuremath{PC}\xspace}

\newcommand{\ER}{\ensuremath{ER}\xspace}

\newcommand{\ekr}{\ensuremath{eKR}\xspace}
\newcommand{\ROM}{\ensuremath{ROM}\xspace}
\newcommand{\VR}{\ensuremath{VR}\xspace}
\newcommand{\DE}{\ensuremath{DE}\xspace}

\newcommand{\metadata}{\textit{METADATA}\xspace}
\newcommand{\gpio}{\texttt{GPIO}\xspace}

\renewcommand{\prover}{{\ensuremath{\sf{\mathcal Prv}}}\xspace} 
\renewcommand{\verifier}{{\ensuremath{\sf{\mathcal Vrf}}}\xspace} 
\newcommand{\dev}{{\ensuremath{\sf{\mathcal Dev}}}\xspace}
\newcommand{\ctrl}{{\ensuremath{\sf{\mathcal Ctrl}}}\xspace} 

\newcommand{\RA}{{\ensuremath{\sf{\mathcal RA}}}\xspace}

\newcommand{\chal}{{\ensuremath{\sf{\mathcal Chal}}}\xspace}
\newcommand{\authorize}{\ensuremath{\mathsf{Authorize}}\xspace}
\renewcommand{\verify}{\ensuremath{\mathsf{Verify}}\xspace}

\newcommand{\hwmon}{\ensuremath{\mathsf{Hardware Monitor}}\xspace}
\newcommand{\atomic}{\ensuremath{\mathsf{atomicExec}}\xspace}
\newcommand{\privsense}{\ensuremath{\mathsf{XSensing}}\xspace}

\renewcommand{\adv}{\ensuremath{\sf{\mathcal Adv}}\xspace}

\newcommand{\reset}{\ensuremath{reset}\xspace}

\newcommand{\fw}{{\ensuremath{\sf{\mathcal S}}}\xspace}
\newcommand{\mem}{{\ensuremath{\sf{\mathcal M}}}\xspace}

\newcommand{\stat}{{\ensuremath{\sf{s}}}\xspace}
\newcommand{\inst}{{\ensuremath{\sf{i}}}\xspace}

\newcommand{\readset}{{\textit{READ}}\xspace}
\newcommand{\writeset}{{\textit{WRITE}}\xspace}
\newcommand{\dmaset}{{\textit{DMA}}\xspace}
\newcommand{\irqset}{{\textit{IRQ}}\xspace}
\newcommand{\resetset}{{\textit{RESET}}\xspace}
\newcommand{\execution}{{\textit{EXEC}}\xspace}

\mathchardef\mhyphen="2D

\usepackage{tikz}

\newtheorem{definition}{Definition}

\newtheorem{theorem}{Theorem}
\newtheorem{construction}{Construction}

\begin{document}

\title{\pfbtext: Protecting Sensed Data from Malicious Sensors with \tt{VERSA}}

\author{\IEEEauthorblockN{Ivan De Oliveira Nunes}
\IEEEauthorblockA{\textit{Rochester Institute of Technology} \\
ivanoliv@mail.rit.edu}
\and
\IEEEauthorblockN{Seoyeon Hwang}
\IEEEauthorblockA{\textit{UC Irvine} \\
seoyh1@uci.edu}
\and
\IEEEauthorblockN{Sashidhar Jakkamsetti}
\IEEEauthorblockA{\textit{UC Irvine} \\
sjakkams@uci.edu}
\and
\IEEEauthorblockN{Gene Tsudik}
\IEEEauthorblockA{\textit{UC Irvine} \\
gene.tsudik@uci.edu}
}

\maketitle

\begin{abstract}
With the growing popularity of the Internet-of-Things (IoT), massive numbers of specialized devices 
are deployed worldwide, in many everyday settings, including homes, offices, vehicles, public spaces, and factories.  
Such devices usually perform sensing and/or actuation. Many of them handle sensitive and personal data. 
If left unprotected, ambient sensing (e.g., of temperature, motion, audio, or video) can leak
very private information. At the same time, some IoT devices use low-end computing platforms with few (or no)
security features.

There are many well-known techniques to secure sensed data, e.g., by authenticating communication end-points, encrypting 
data before transmission, and obfuscating traffic patterns. Such techniques protect sensed data from external adversaries 
while assuming that the sensing device itself is secure. Meanwhile, both the scale and frequency of IoT-focused attacks 
are growing. This prompts a natural question: \textit{how to protect sensed data even if all software on the device is compromised?}
Ideally, in order to achieve this, sensed data must be protected from its genesis, i.e., from the time when a physical analog 
quantity is converted into its digital counterpart and becomes accessible to software. We refer to this property as \pfb: \pfbtext.

In this work, we formalize \pfb and design \ppstextunderlined (\pps) -- a provably secure and formally verified
architecture guaranteeing that only correct execution of expected and explicitly authorized software can access and 
manipulate sensing interfaces, specifically, General Purpose Input/Output (GPIO), which is the usual boundary between 
analog and digital worlds on IoT devices. This guarantee is obtained with minimal hardware support 
and holds even if all device software is compromised. \pps ensures that malware can neither gain 
access to sensed data on the GPIO-mapped memory nor obtain any trace thereof.  
\pps is formally verified and its open-sourced implementation targets resource-constrained IoT edge devices, commonly 
used for sensing. Experimental results show that \pfb is both achievable and affordable for such devices.
\end{abstract}

\section{Introduction}\label{sec:intro}
The impact and importance of embedded (aka IoT or ``smart'') devices is hard to overestimate.
They are increasingly popular and becoming pervasive in many settings: from 
homes and offices to public spaces and industrial facilities. Not surprisingly, they also represent increasingly
attractive attack targets for exploits and malware. In particular, low-end (cheap, small, and simple) micro-controller units 
(MCUs) are designed with strict cost, size, and energy limitations. Thus, it is hard to offer 
any concrete guarantees for tasks performed by these MCUs, due to their lack of sophisticated 
security and privacy features, compared to higher-end computing devices, such as smartphones or 
general-purpose IoT controllers, e.g., Amazon Echo or Google Nest. As MCUs increasingly permeate private spaces, 
exploits that abuse their sensing capabilities to obtain sensitive data represent a significant privacy threat.

Over the past decade, the IoT privacy issues have been recognized and explored by the research 
community~\cite{priv_iot0,priv_iot1,priv_iot2,priv_iot3,priv_iot4}. Many techniques (e.g.,
\cite{neto2016aot,kumar2019jedi}) were developed to secure sensor data from active attacks 
that impersonate users, IoT back-ends, or servers.  Another research direction focused on 
protecting private data from passive in-network observers that intercept traffic 
\cite{trimananda2020packet,apthorpe2018keeping,apthorpe2017closing,apthorpe2017smart} or 
perform traffic analysis based on unprotected packet headers and other 
metadata, e.g., sizes, timings, and frequencies. However, security of sensor data {\bf on the device} 
which originates that data has not been investigated. We consider this to be a crucial issue, 
since all software on the device can be compromised
and leak (exfiltrate) sensed data.
Whereas, aforementioned techniques assume that sensing device runs the expected {\bf benign} software.

We claim that in order to solve this problem, privacy of sensed data must be ensured ``from birth''.
This corresponds to two requirements: (1) access to sensing interfaces must be strictly controlled, 
such that only authorized code is allowed to read data, (2) sensed data must be protected 
as soon as it is converted to digital form. Even the simplest devices (e.g., motion sensors, 
thermostats, and smart plugs) should be protected since prior 
work~\cite{cheng2017homespy,narain2016inferring,sukor2019hybrid,jin2017virtual} amply
demonstrates that private 
-- and even safety-critical -- information can be inferred from sensed data. It is also well-known
that even simple low-end IoT devices are subject to malware attacks. This prompts a natural question: 
\emph{Can privacy of sensed data be guaranteed if the device software is compromised?}
We refer to this guarantee as {\small \pfbtext} (\pfb).

Some previous results considered potential software compromise in low-end devices and proposed methods
to enable security services, such as remote verification of device software state (remote attestation) 
\cite{smart,sancus,vrasedp,simple,tytan,trustlite}, proofs of remote software execution~\cite{apex}, 
control- \& data-flow attestation~\cite{litehax,cflat,lofat,atrium,oat,tinycfa,dialed}, as well as 
proofs of remote software updates, memory erasure, and system reset~\cite{pure,verify_and_revive,asokan2018assured}.

Regardless of their specifics, such techniques only \underline{detect} of violations or compromises {\bf after the fact}.
In the context of \pfb, that is too late since leakage of private sensed data likely already occurred.
Notably, SANCUS~\cite{sancus} specifically discusses the problem of access control to sensor 
peripherals (e.g., GPIO) and proposes attestation of software accessing (or controlling access to) these peripherals. 
However, this only allows detection of compromised peripheral-accessing software and does not
prevent illegal peripheral access.

To bridge this gap and obtain \pfb, we construct the \ppstextunderlined (\pps) architecture. It provably
prevents leakage of private sensor data even when the underlying device is software-compromised.
At a high level, \pps combines three key features: (1) \emph{Mandatory Sensing Operation Authorization}, 
(2) \emph{Atomic Sensing Operation Execution}, and (3) \emph{Data Erasure on Boot} (see Section~\ref{sec:overview}). 
To attain these features, \pps implements a minimal and formally verified hardware monitor that runs independently from 
(and in parallel with) the main CPU, without modifying the CPU core. We show that \pps is an efficient and
inexpensive means of guaranteeing \pfb. 

This work makes the following contributions:
\begin{compactitem}
	\item Formulates \pfb with a high-level specification of requirements, followed by a 
	game-based formal definition of the \pfb goal. 
 	\item Constructs \pps, an architecture that guarantees \pfb.
 	\item Implements and deploys \pps on a commodity low-end MCU, which demonstrates its cost-effectiveness and practicality. 
 	\item Formally verifies \pps implementation and proves security of the overall construction, hence 
        obtaining provable security at both architectural and implementation levels.  \pps implementation 
        and its computer proofs are publicly available in \cite{pfb-repo}.
\end{compactitem}

\section{Preliminaries}\label{sec:BG}
\begin{figure}
	\centering
	\includegraphics[width=1\columnwidth]{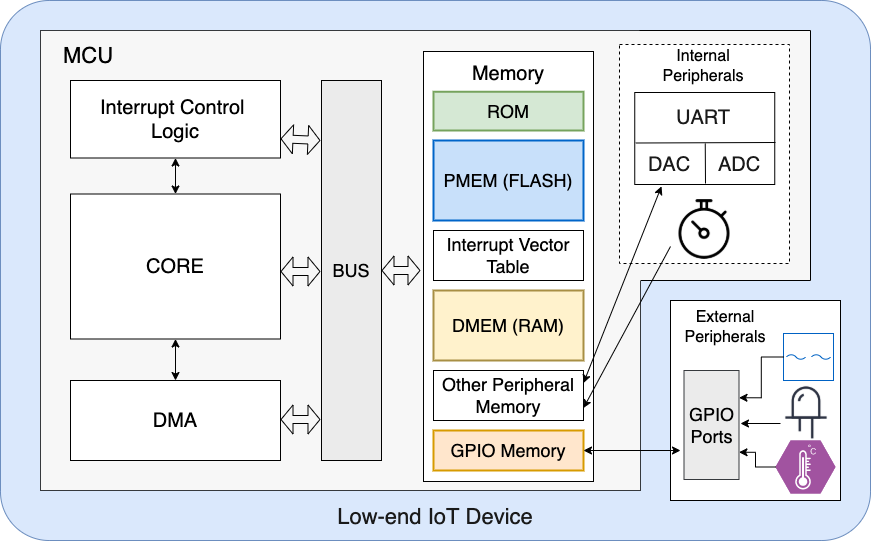}
	\vspace{-0.5em}
	\caption{\footnotesize System Architecture of an MCU-based IoT Device}
	\vspace{-1em}
	\label{fig:mcu_arch}
\end{figure}

\subsection{Scope \& MCU-based devices}\label{subsec:scope}
This work focuses on low-end CPS/IoT/smart devices with low computing power and meager resources.
These are some of the smallest and weakest devices based on low-power single-core MCUs with only a few kilobytes 
(KB) of program and data memory.  Two prominent examples
are Atmel AVR ATmega and TI MSP430: $8$- and $16$-bit CPUs, respectively, typically running at $1$-$16$MHz clock frequencies, 
with $\approx$ 64KB of addressable memory. 

Figure \ref{fig:mcu_arch} illustrates a generic architecture representing such MCUs. The CPU core and the Direct-Memory Access 
(DMA) controller access memory through a bus.\footnote{DMA is a hardware controller that can read/write to memory in parallel with the CPU.}
Memory can be divided into 5 logical regions: (1) Read-only memory (ROM), if present, stores critical software such as a bootloader, 
burnt into the device at manufacture time and not modifiable thereafter; (2) program memory (PMEM), usually realized as flash, is 
non-volatile and stores program instructions; (3) interrupt vector table (also in flash and often considered as part of PMEM), 
stores interrupt configurations; (4) data memory (DMEM), usually implemented with DRAM, is volatile and used to store program 
execution state, i.e., its stack and heap; and, (5) peripheral memory region (also in DRAM and often considered as a part of 
DMEM), contains memory-mapped I/O interfaces, i.e., addresses in the memory layout that are mapped to hardware components,
e.g., timers, UART, and GPIO. In particular, GPIO are peripheral memory addresses hardwired to physical ports that  
interface with external circuits, e.g., analog sensors/circuits.

We note that small MCUs usually come in one of two memory architectures: Harvard and von Neumann. 
The former isolates PMEM and DMEM by maintaining two different buses and address spaces, while the latter keeps 
both PMEM and DMEM in the same address space and accessible via a single bus. 

Low-end MCUs execute instructions in place, i.e., directly from flash memory. They have neither memory management units 
(MMUs) to support virtualization/isolation, nor memory protection units (MPUs). Therefore, privilege levels and isolation used 
in higher-end devices and generic enclaved execution systems 
(e.g., Intel SGX~\cite{sgx} or MIT SANCTUM~\cite{sanctum}) are not applicable. 

We believe that a \pfb-agile architecture that is sufficiently inexpensive and efficient for such low-end devices 
can be later adapted to more powerful devices. Whereas, going in the opposite direction is more challenging. 
Furthermore, simpler devices are easier to model and reason about formally. Thus, we believe that they represent a 
natural starting point for the design and verification of a \pfb-agile architecture. 
To this end, our prototype implementation of \pps is integrated with MSP430, due in part to public availability 
of an open-source MSP430 hardware design from OpenCores \cite{openmsp430}.\footnote{Nevertheless, 
the generic machine model and methodology of \pps are  applicable to other low-end MCUs of the same class, 
e.g., Atmel AVR ATmega.}

\subsection{\gpio \& MCU Sensing}\label{sec:gpio_bg}
A GPIO port is a set of GPIO pins arranged and controlled together, as a group. The MCU-addressable
memory for a GPIO port is physically mapped (hard-wired) to physical ports that can be connected to a variety of 
external circuits, such as analog sensors and actuators, as shown in Figure~\ref{fig:mcu_arch}.
Each GPIO pin can be set to function as either an input or output, hence called "general purpose".
Input signals produced by external circuits can be obtained by the MCU software by reading from GPIO-mapped memory.
Similarly, egress electric signals (high or low voltage) can be generated by the MCU software by writing (logical $1$ or $0$) 
to GPIO-mapped memory.

\textit{\textbf{Remark:}} ``GPIO-mapped memory'' includes the set of all software-readable memory regions connected to 
external sensors. In some cases, this set may even include multiple physical memory regions for a single physical pin. 
For instance, if a given GPIO pin is also equipped with an Analog-to-Digital Converter (ADC), a GPIO input could be 
reflected on different memory regions depending on whether the ADC is active or inactive. All such regions are considered 
``GPIO-mapped memory'' and we refer to it simply as \gpio.
Using this definition, in order to access sensor data, software running on the MCU must read from \gpio. 

We also note that various applications require different sensor regimes~\cite{WSNbook2010}: event-driven, 
periodic, and on-demand.  Event-driven sensors report sensed data when a trigger event occurs, while 
periodic sensors report sensor data at fixed time intervals. On-demand (or query-driven) sensors report sensor 
data whenever requested by an external entity.  Although we initially consider on-demand sensing, as discussed 
in Section~\ref{sec:overview}, the proposed design is applicable to other regimes.

\subsection{\new{Remote Attestation \& \vrased}} \label{subsec:BG-vrased}
Remote Attestation (\RA) allows a trusted entity (verifier = \verifier) to remotely measure current memory contents (e.g., software binaries) 
of an untrusted embedded device (prover = \prover). \RA is usually realized as a challenge-response protocol:
\begin{compactenum}
 \item \verifier sends an attestation request containing a challenge (\chal) to \prover.  
 \item \prover receives the request  and computes an authenticated integrity check 
 	over its memory and \chal. The memory region can be either pre-defined or explicitly specified in the attestation request.
 \item \prover returns the result to \verifier.
 \item \verifier verifies the result and decides if it corresponds to a valid \prover state.
\end{compactenum}
\new{
\vrased~\cite{vrasedp} is a verified hybrid (hardware/software) \RA architecture for for low-end MCUs. 
It comprises a set of (individually) verified hardware and software sub-modules; their composition 
provably satisfies formal definitions of \RA soundness and security.
\vrased software component implements the authenticated integrity function computed over a given 
``Attested Region'' (AR) of \prover's memory.
\vrased hardware component assures that its software counterpart executes securely and that no 
function of the secret key is ever leaked.
In short, \RA soundness states that the integrity measurement must accurately reflect a snapshot of \prover's memory in AR, 
disallowing any modifications to AR during the actual measurement. \RA security defines that the measurement must be 
unforgeable, implying protection of secret key \attkey used for the measurement.
}

In order to prevent DoS attacks on \prover, the RA protocol may involve authentication of the attestation request, before
\prover performs attestation. If this feature is used, an authentication token must accompany every attestation 
request.\footnote{By saying ``this feature is used'', we mean that its usage (or lack thereof) is fixed at the granularity 
of a \verifier-\prover setting, and not per single \RA instance.}
For example, in \vrased, \verifier computes this token as an HMAC over \chal, using \attkey. Since \attkey is only known 
to \prover and \verifier, this token is unforgeable. To prevent replays, \chal is a monotonically increasing counter, and 
the latest \chal used to successfully authenticate \verifier is stored by \prover in persistent and protected memory. 
In each attestation request, incoming \chal must be greater that the stored value. 
Once an attestation request is successfully authenticated, the stored value is updated accordingly. 

\new{ 
\vrased software component is stored in ROM and realized with a formally verified HMAC implementation
from the HACL* cryptographic library \cite{hacl}, which is used to compute: 
$
H=HMAC(KDF(\attkey, \chal), AR),
$
where $KDF(\attkey, \chal)$ is a one-time key derived from the received \chal and \attkey using a key derivation function.
}

As discussed later in Section~\ref{sec:design}, in \pps, \vrased is used as a means of authorizing a binary to access \gpio.

\subsection{LTL, Model Checking, \& Verification}
Our verification and proof methodologies are in-line with prior work on the design and verification of
security architectures proving code integrity and execution properties for the same class of MCUs~\cite{vrasedp, apex, rata, tarot}. 
However, to the best of our knowledge, no prior work tackled formal models and 
definitions, or designed services, for guaranteed sensed data privacy. 
This section overviews our verification and proof methodologies that allow us to later show that 
\pps achieves required \pfb properties and end-goals.

Computer-aided formal verification typically involves three steps. First, the system of interest (e.g., hardware, 
software, or communication protocol) is described using a formal model,  e.g., a Finite State Machine (FSM). 
Second, properties that the model should satisfy are formally specified. Third, the system model is checked against 
formally specified properties to guarantee that the system retains them. This can be done via Theorem 
Proving~\cite{loveland2016automated} or Model Checking~\cite{clarke2018model}.
We use the latter to verify the implementation of system sub-modules, and the former to 
prove new properties derived from the combination (conjunction) of machine model axioms and 
sub-properties that were proved for the implementation of individual sub-modules.

In one instantiation of model checking, properties are specified as \textit{formulae} using Linear Temporal Logic (LTL) 
and system models are represented as FSMs. Hence, a system is represented by a triple: $(\sigma, \sigma_0, T)$, 
where $\sigma$ is the finite set of states, $\sigma_0 \subseteq \sigma$ is the set of possible initial states, and 
$T \subseteq \sigma \times \sigma$ is the transition relation set, which describes the set of states that can be reached 
in a single step from each state. Such usage of LTL allows representing a system behavior over time.

Our verification strategy benefits from the popular model checker NuSMV~\cite{nusmv}, which can 
verify generic hardware or software models. For digital hardware described at Register Transfer Level (RTL) -- which is the 
case in this work -- conversion from Hardware Description Language (HDL) to NuSMV models is simple. Furthermore, 
it can be automated~\cite{irfan2016verilog2smv} as the standard RTL design already relies on describing 
hardware as FSMs. LTL specifications are particularly useful for verifying sequential systems. In addition to propositional 
connectives, such as conjunction ($\land$), disjunction ($\lor$), negation ($\neg$), and implication ($\rightarrow$), 
LTL extends propositional logic with {\bf temporal quantifiers}, thus enabling sequential reasoning. In this paper, 
we are interested in the following LTL quantifiers:

\vspace*{0.2cm}
\fbox{\begin{minipage}[c]{0.9\linewidth} \footnotesize	
\noindent$\bullet$ \textbf{X}$\phi$ -- ne\underline{X}t $\phi$: holds if $\phi$ is true at the next system state.\\
$\bullet$ \textbf{G}$\phi$ -- \underline{G}lobally $\phi$: holds if for all future states $\phi$ is true.\\
$\bullet$ $\phi$ \textbf{U} $\psi$ -- $\phi$ \underline{U}ntil $\psi$: holds if there is a future state 
	where $\psi$ holds and $\phi$ holds for all states prior to that.\\
$\bullet$ $\phi$ \textbf{W} $\psi$ -- $\phi$ \underline{W}eak until $\psi$: holds if, assuming a future state where $\psi$ 
	holds, $\phi$ holds for all states prior to that. If $\psi$ never becomes true, $\phi$ must hold forever. 
	Or, more formally: $\phi$ \textbf{W} $\psi \equiv (\phi$ \textbf{U} $\psi) \lor$ \textbf{G}$(\phi)$.\\
$\bullet$ $\phi$ \textbf{B} $\psi$ -- $\phi$ \underline{B}efore $\psi$: holds if the existence of state where $\psi$ 
 	holds implies the existence of at least one earlier state where $\phi$ holds. Equivalently: 
	$\phi$ \textbf{B} $\psi \equiv \neg(\neg\phi$ \textbf{U} $\psi)$.
\end{minipage}}
\vspace*{0.2cm}

NuSMV works by exhaustively enumerating all possible states of a given system FSM and by checking each state 
against LTL specifications. If any desired specification is found not to hold for specific states 
(or transitions between states), the model checker provides a trace that leads to the erroneous state, 
which helps correct the implementation accordingly. As a consequence of exhaustive enumeration, proofs for complex 
systems that involve complex properties often do not scale due to the so-called ``state explosion'' problem.
To cope with it, our verification approach is to specify smaller LTL sub-properties separately and verify each 
respective hardware sub-module for compliance. In this process, our verification pipeline automatically 
converts digital hardware, described at RTL using Verilog, to Symbolic Model Verifier 
(SMV)~\cite{smv} FSMs using Verilog2SMV~\cite{irfan2016verilog2smv}.  The SMV representation is then fed to 
NuSMV for verification. Then, the composition of LTL sub-properties (verified in the model-checking 
phase) is proven to achieve a desired end-to-end implementation goal, also specified in LTL. This step uses an 
LTL theorem prover~\cite{spot}. 

In our case, we show that the end-to-end goal of \pps, in composition with \vrased, 
is sufficient to achieve \pfb via cryptographic reduction from the formal security definition of \vrased.
These steps are discussed in detail in Section~\ref{sec:security_impl}.

\section{\pps Overview}\label{sec:overview}
\pps involves two entities: a trusted remote controller (\ctrl) and a device (\dev).
We expect \ctrl to be a relatively powerful computing entity, e.g., a home gateway, a backend server or even a smartphone. 
\pps protects sensed data on \dev by keeping it (and any function thereof) confidential. This implies: (1) controlling 
GPIO access by blocking attempted reads by unauthorized software, and (2) keeping execution traces 
(i.e., data allocated by \gpio-authorized software) confidential.
Therefore, access to \gpio is barred by default. \gpio is unlocked only for benign binaries that are pre-authorized by \ctrl.
Whenever a binary is deemed to be authorized on \dev, \pps creates for it an ephemeral isolated execution 
environment and permits its one-time execution.
This isolated environment lasts until execution ends, which corresponds to reaching the legal exit point of the 
authorized binary. Therefore, by including a clean-up routine immediately before the legal exit, we can assure that
all execution traces,  including all sensitive information, are erased. Any attempt to interrupt, or tamper with, 
isolated execution causes an immediate system-wide reset, which erases all data traces.

\new{
We use the term ``Sensing Operation'', denoted by \fw, to refer to a self-contained and logically independent binary 
(e.g., a function) that is responsible for processing data obtained through one or more reads from \gpio. 
}

\pps achieves \pfb via three key features: 

\noindent\emph{{\bf [A]} Mandatory Sensing Operation Authorization} requires explicit authorization issued by \ctrl before any 
\dev software reads from \gpio. Recall that access to \gpio is blocked by default.
\new{
Each authorization token (\atoken) coming from \ctrl allows one execution of a specific sensing operation \fw, 
although a single execution of \fw can implement several \gpio reads.
}
\atoken has the following properties:
\begin{compactenum}
\item It can be authenticated by \dev as having been issued by \ctrl; this includes freshness;
\item It grants privileges {\bf only to a specific} \fw to access \gpio during its execution; and
\item It can only be used once.
\end{compactenum}
\new{
\ctrl can authorize multiple executions of \fw by issuing a batch of tokens, i.e.,  $\atoken_1,...,\atoken_n$, for up to
$n$ executions of \fw.} Although supporting multiple tokens is unnecessary for on-demand sensing, it might be 
useful for periodic or event-driven sensing regimes discussed in Section~\ref{sec:gpio_bg}.

\noindent\emph{{\bf [B]} Atomic Sensing Operation Execution} ensures that, once authorized by \ctrl, \fw is executed 
with the following requirements:
\begin{compactenum}
\item \fw execution starts from its legal entry point (first instruction) and runs until its legal exit point (last instruction). 
This assumes a single pair of entry-exit points;
\item \fw execution can not be interrupted and its intermediate  results cannot be accessed by
external means, e.g., via DMA controllers; and 
\item An immediate MCU reset is triggered if either (1) or (2) above is violated.
\end{compactenum}

\begin{figure}
	\centering
	\includegraphics[width=1\columnwidth]{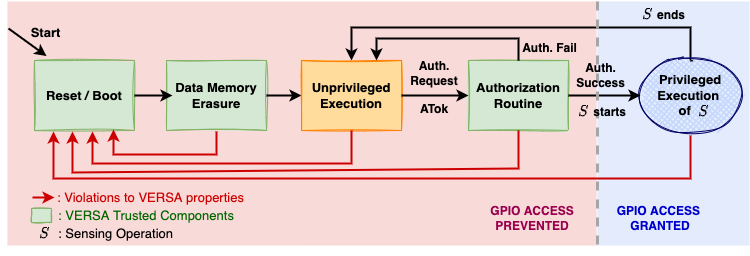}
	\vspace{-0.5em}
	\caption{\footnotesize \new{MCU execution workflow with \pps.}}
	\label{fig:workflow}
\end{figure}

\noindent\new{\emph{{\bf [C]} Data Erasure on Reset/Boot} works with \emph{\bf [B]} to 
guarantee that, sensed data (or any function thereof) obtained during \fw execution is not leaked due to errors or 
violations of security properties, which cause MCU reset per item (3) above. This feature must guarantee that 
all values that remain in RAM after a hard reset and the subsequent boot process, are erased before any 
unprivileged software can run. While some architectures already provide memory erasure on boot, for those MCUs that 
do not do so, it can be obtained by calling a secure RAM erasure function at boot time, e.g., as a part of a ROM-resident 
bootloader code. Appendix~\ref{apdx:erasure} discusses this further.  
}

At a high level, correct implementation of aforementioned three features suffices to obtain \pfb, because:
\begin{compactitem}
	\item Any compromised/modified binary can not access \gpio since it has no authorization from \ctrl.
	\item Any authorized binary \fw must be invoked properly and run atomically, from its first, and until its last, instruction.
	\item Since \fw is invoked properly, intended behavior of \fw is preserved. Code reuse attacks are not possible, unless they
	occure as a result of bugs in \fw implementation itself. \ctrl can always check for such bugs in \fw prior to authorization;
	see Section~\ref{sec:adv_model}.
	\item \fw runs uninterrupted, meaning that it can erase all traces of its own execution from the stack
	before passing control to unprivileged applications. This guarantees that no sensor data remains in 
	memory when \fw terminates.
	\item \pps assures that any violation of aforementioned requirements causes an MCU reset, triggering erasure of 
	all data memory. Therefore, malware that attempts to interrupt \fw before completion, or tamper with 
	\fw execution integrity, will cause all data used by \fw to be erased.
\end{compactitem}
\vspace{0.5mm}

\noindent{\bf Support for Output Encryption:}
\fw might process and use sensor data locally as part of its own execution, or generate some output that needs to be returned to 
\ctrl. In the latter case, encryption of \fw output is necessary. For this reason, \pps supports the generation of a fresh  
key derived from \atoken (thus implicitly shared between \ctrl and \fw). This key is only accessible to \fw during authorized execution. 
Hence, \fw can encrypt any data to be exported with this key and ensure that encrypted results can only be decrypted by \ctrl.

Since we assume that the encryption function is part of \fw, it cannot be interrupted (or tampered with) by any unprivileged 
software or external means. Importantly, the encryption key is only accessible to \fw (similar to \gpio) and shielded from 
all other software. Furthermore, the choice of the encryption algorithm is left up to the specific \fw implementation. 

Figure ~\ref{fig:workflow} illustrates MCU execution workflow discussed in this section. 

\begin{figure*}
	\begin{mdframed}
		\begin{definition}[{\small MCU Execution Model}]\label{def:model}~\\
			\footnotesize
			\vspace*{-0.6em}
			
			\textbf{1 -- Execution} is modeled as a sequence of MCU states $\texttt{S}:=\{\stat_0, ...,\stat_m\}$ and a sequence of instructions 
			$\texttt{I}:=\{\inst_0,...,\inst_n\}$. Since the \underline{next} MCU state and the \underline{next} instruction to be executed are 
			determined by the \underline{current} MCU state and the \underline{current} instruction being executed,
			these discrete transitions are denoted as shown in the following example:
			\vspace*{-0.2em}
			\begin{equation*}
			\boldsymbol{(\stat_{1}, \inst_{j}) \leftarrow \textbf{\execution}(\stat_0,\inst_0)};\quad
			\boldsymbol{(\stat_{2}, \inst_{k}) \leftarrow \textbf{\execution}(\stat_1,\inst_j)};\quad
			\textbf{~...}\quad
			\boldsymbol{(\stat_{m}, \perp) \leftarrow \textbf{\execution}(\stat_{m-1},\inst_l)}
			\end{equation*}
			The sequence $\texttt{I}$ represents the physical order of instructions in memory, which is 
			not necessarily the order of their execution. 
			The next instruction and state are also affected by current external inputs, current data-memory values, 
			and current hardware events, e.g., interrupts or resets, which are modeled as properties of each execution 
			state in $\texttt{S}$. The MCU always starts execution (at boot or after a reset) from state $\stat_0$ 
			and initial instruction $i_0$. \execution produces $\perp$ as the next instruction if there is no instructions left to execute.
			
			\vspace{2mm}
			\textbf{2 -- State Properties as Sets:} sets are used to model relevant execution properties and characterize 
			effects/actions occurring within a given state $\stat_t$. We are particularly interested in the behaviors
			corresponding to the following sets:
			\begin{compactenum}
				\item \textbf{\readset: } all states produced by the execution of an instruction $\inst$ 
				that reads the value from memory to a register.
				\item \textbf{\writeset: }all states produced by the execution of an instruction $\inst$ 
				that writes the value from a register to memory. 
				\item \textbf{$\dmaset^R$: } all states produced as a result of DMA reading from memory.
				\item \textbf{$\dmaset^W$: } all states $\stat_t$ produced as a result of DMA writing to memory.
				\item \textbf{\irqset: } all states $\stat_t$ where an interrupt is triggered.
				\item \textbf{\resetset: } all states $\stat_t$ wherein an MCU reset is triggered.
			\end{compactenum}
			\vspace{0.3mm}
			Note that these sets are not disjoint, i.e., $\stat_t$ can belong to multiple sets. 
			Also, the aforementioned sets do not aim to model 
			all possible MCU behaviors, but only the ones relevant to \pfb.
			Finally, we further subdivide sets that model memory access into subsets relating to memory regions of interest. 
			For example, considering a contiguous memory region $\mem = [\mem_{min},\mem_{max}] $, $\readset_\mem$ 
			is a subset of \readset containing only the states produced through \execution of instructions that read from 
			the memory region $\mem$. We use the same notation to refer to other subsets, e.g., $\writeset_\mem$, 
			$\dmaset^R_\mem$, and $\dmaset^W_\mem$.
		\end{definition}
	\end{mdframed}
	\vspace*{-0.5em}
	\begin{mdframed}
		\linespread{0.4}
		\begin{definition}[{\small Hardware Model}]\label{def:hw_model}~\\
			\footnotesize
			
			\mem denotes a contiguous memory region within addresses $\mem_{min}$ and $\mem_{max}$ 
			in physical memory of \dev, i.e., $\mem := [\mem_{min}, \mem_{max}]$.
			
			\stat represents the system execution state at a given CPU cycle.\\ \\
			
			{\bf Program counter \& instruction execution:}
			\vspace*{-0.8em}
			\begin{align}
			\begin{split}\label{ltl:cpu_pc}
			{\bf G:}\{[{\bf X}(\stat) \leftarrow \execution(\stat,\inst_k) \land \inst_k \in \mem)] \rightarrow (PC \in \mem)\}
			\end{split}
			\end{align}
			
			{\bf Memory Reads/Writes:} 
			\vspace*{-0.8em}
			\begin{align}
			\begin{split}\label{ltl:cpu_mem_read}
			{\bf G:}\{{\bf X}(\stat) \in \readset_\mem \rightarrow (R_{en} \land D_{addr} \in \mem)\}
			\end{split}
			\end{align}

			\vspace*{-1.2em}
			\begin{align}
			\begin{split}\label{ltl:cpu_mem_write}
			{\bf G:}\{{\bf X}(\stat) \in \writeset_\mem \rightarrow (W_{en} \land D_{addr} \in \mem)\}
			\end{split}
			\end{align}

			\vspace*{-1.2em}
			\begin{align}
			\begin{split}\label{ltl:DMA_mem_read_write}
			{\bf G:}\{({\bf X}(\stat) \in \dmaset^R_\mem \lor {\bf X}(\stat) \in \dmaset^W_\mem)\rightarrow (DMA_{en} \land DMA_{addr} \in \mem)\}
			\end{split}
			\end{align}
			%
			{\bf Interrupts (\irq) and Resets:}
			\vspace*{-0.8em}
			\begin{align}\label{ltl:trigger_interrupt}
			\begin{split}
			{\bf G:}\{\stat \in \irqset \leftrightarrow irq\}
			\end{split}
			\end{align}
			\vspace*{-1.2em}
			\begin{align}\label{ltl:trigger_reset}
			\begin{split}
			{\bf G:}\{\stat \in \resetset \leftrightarrow reset\}
			\end{split}
			\end{align}
		\end{definition}
	\end{mdframed}
	\vspace{-1.5em}
\end{figure*}

\section{MCU Machine Model} \label{sec:machine_model}
\subsection{Execution Model}\label{sec:exec_model}
To enable formal specification of \pfb guarantees, we formulate the MCU execution model in Definition~\ref{def:model}.
It represents MCU operation as a discrete sequence of MCU states, each corresponding to one clock cycle -- the 
smallest unit of time in the system. We say that the subsequent MCU state is defined based on the current MCU state
(which includes current values in memory/registers, as well as any hardware signals and effects, such as
external inputs, actions by DMA controller(s), and interrupts) and 
the current instruction being executed by the CPU core. Similarly, the instruction to be executed 
in the next state is determined by the current state and the current instruction being executed.

For example, an arithmetic instruction (e.g., \texttt{add} or \texttt{mult}) causes the program counter (\pc) to point to 
the subsequent address in physical memory. However, an interrupt (which is a consequence of the current MCU state) 
may occur and deviate the normal execution flow. Alternatively, a branching instruction may be executed and 
cause \pc to jump to some arbitrary instruction that is not necessarily located at the subsequent position 
in the MCU flash memory.

In order to reason about events during the MCU operation, we say that each MCU state can belong to one or 
more sets. Belonging to a given set implies that the state has a given property of interest. Definition~\ref{def:model} 
introduces six sets of interest, representing states in which memory is read/written by 
CPU or DMA, as well as states in which an interrupt or reset occurs.

\subsection{Hardware Signals}\label{sec:hw_model}
We now formalize the effects of execution, modeled in Definition~\ref{def:model}, 
to the values of concrete hardware signals that can be monitored by \pps hardware 
in order to attain \pfb guarantees. Informally, we model the following simple axioms:

\begin{compactitem}
	\item[\textbf{[A1] PC:}] contains the memory address containing 
	the instruction being executed at a given cycle.
	\item[\textbf{[A2] CPU Memory Access:}] Whenever memory is read or written, a data-address 
	signal (\daddr) contains the address of the corresponding memory location. A data read-enable bit (\ren) 
	must be set for a read access and a data write-enable bit (\wen) must be set for a write access.
	\item[\textbf{[A3] DMA:}] Whenever a DMA controller attempts to access 
	the main memory, a DMA-address signal (\dmaaddr) contains the address of the accessed memory 
	location and a DMA-enable bit (\dmaen) must be set. 
	\item[\textbf{[A4] Interrupts:}] When hardware interrupts or software interrupts happen, the \irq~signal is set.
	\item[\textbf{[A5] MCU reset:}] At the end of a successful reset routine, all registers (including \pc) are 
	set to zero before restarting software execution. The reset handling routine cannot be modified, 
	as resets are handled by MCU in hardware. When a reset happens, the corresponding \reset signal is set. 
	The same signal is also set when the MCU initializes for the first time.
\end{compactitem}
\noindent This model strictly adheres to MCU specifications, assumed to be correctly implemented  
by the underlying MCU core.

Definition~\ref{def:hw_model} presents formal specifications for  aforementioned axioms in LTL. Instead of  
explicitly quantifying time, LTL embeds time within the logic by using temporal quantifiers (see
Section~\ref{sec:BG}). 
Hence, rather than referring to execution states 
using temporal variables (i.e., state $t$, state $t+1$, state $t+2$), 
a single variable (\stat) 
and LTL quantifiers suffice to specify, e.g.,  ``current'', ``next'', ``future'' system states (\stat). 
For this part of the model, we are mostly interested in: (1) describing MCU state at the next CPU cycle 
(${\bf X}(\stat)$) as a function of the MCU state at the current CPU cycle (\stat), and (2) describing which particular MCU signals must be triggered in order for ${\bf X}(\stat)$ to be in each of the sets defined in Definition~\ref{def:model}.

\new{
LTL statements in Definition~\ref{def:hw_model} formally model axioms {\bf [A1]-[A5]}, i.e., 
the subset of MCU behavior that is relevant to, and sufficient for formally verifying, \pps.
LTL (\ref{ltl:cpu_pc}) models {\bf [A1]}, (\ref{ltl:cpu_mem_read}) and (\ref{ltl:cpu_mem_write}) model {\bf [A2]}, 
and each (\ref{ltl:DMA_mem_read_write}), (\ref{ltl:trigger_interrupt}), and (\ref{ltl:trigger_reset}) models {\bf [A3]}, 
{\bf [A4]}, and {\bf [A5]}, respectively.
}

\section{\pfb Definitions}\label{sec:definitions}
\begin{figure*}
	\begin{mdframed}
		\begin{definition}[Syntax: \pfb scheme]\label{def:pfb}~\\
			\footnotesize
			
			A \pfbtext (\pfb) scheme is a tuple of algorithms $[\authorize, \verify, \privsense]$:
			\begin{compactenum}
				\item{$\authorize^{\ctrl}(\fw, \cdots)$:} an algorithm executed by \ctrl taking as \underline{input} at least one 
				executable $\fw$ and \underline{producing} 
				at least one authorization token \atoken which can be sent to \dev to authorize one execution of \fw with 
				access to \gpio.\\ \vspace*{-0.7em}
				\item{$\verify^{\dev}(\fw, \atoken, \cdots)$:} an algorithm (with possible hardware-support), executed by \dev, 
				that takes as \underline{input} \fw and \atoken. It uses \atoken to  
				check whether \fw is pre-authorized by \ctrl and outputs $\top$ if verification succeeds, and $\perp$ otherwise.\\ \vspace*{-0.7em}
				\item{$\privsense^{\dev}(\fw, \cdots)$:} an algorithm (with possible hardware-support) 
				that executes \fw in \dev, producing a sequence of states $E:=\{\stat_0,..., \stat_m\}$. It returns
				$\top$, if sensing successfully occurs during \fw execution, i.e., $\exists 
				\stat \in E$ such that $(\stat \in \readset_{\gpio}) \land (\stat \notin \resetset)$); it returns $\perp$, otherwise.\\ \vspace*{-0.7em}
				\item[\textbf{Remark:}] In the parameter list, ($\cdots$) means that additional/optional parameters might be 
				included depending on the specific \pfb construction.
			\end{compactenum}
		\end{definition}
	\end{mdframed}
	\vspace*{-0.5em}
	\begin{mdframed}
		\vspace*{-0.5em}
		\begin{definition}[\pfb Game-based Definition]\label{def:pfb_sec_def}~\\
			\footnotesize
			\vspace*{-0.5em}
			
			\textbf{\ref{def:pfb_sec_def}.1 Auxiliary Notation \& Predicate(s):}
			\begin{compactitem}
				\item Let \texttt{\attkey} be a secret string of bit-size $|\attkey|$; and $\lambda$ be the security 
				parameter, determined by $|\attkey|$, i.e., $\lambda = \Theta(|\attkey|)$;
				\item Let \atomic be a predicate evaluated on some sequence of states $\texttt{S}$ and some 
				software -- i.e., some sequence of instructions $\texttt{I}$. 
				\begin{compactitem}
					\item $\atomic(\texttt{S}:=\{\stat_1, ..., \stat_m\},\texttt{I}:=\{\inst_0, ..., \inst_n\}) \equiv \top$ if and only if the 
					following hold; otherwise, $\atomic(\texttt{S},\texttt{I}) \equiv \perp$.
					\begin{compactenum}
						\item \textbf{Legal Entry Instruction:} The first execution state $\stat_1$ in $\texttt{S}$ is produced by the execution of the first instruction $\inst_0$ in $\texttt{I}$. \\i.e., $(\stat_1 \leftarrow \execution(\inst_{0},\stat{*})) ~\lor~ (\stat_1 \in \resetset)$, where $\stat{*}$ is any state prior to $\stat_1$.
						
						\item \textbf{Legal Exit Instruction:} The last execution state $\stat_m$ in $\texttt{S}$ is produced by the execution of the last instruction $\inst_n$ in $\texttt{I}$. \\i.e., $(\stat_m \leftarrow \execution(\inst_{n},\stat_{m-1})) ~\lor~ (\stat_m \in \resetset)$.
						
						\item \textbf{Self-Contained Execution:} For all $\stat_j$ in $\texttt{S}$, $\stat_j$ is produced by the execution of an instruction $\inst_k$ in $\texttt{I}$, for some $k$. \\i.e., $(\stat_j \leftarrow \execution(\inst_k,\stat_{j-1})) ~\lor~ (\stat_j \in \resetset)$, for some $\inst_k \in \texttt{I}$.
						
						\item \textbf{No Interrupts, No DMA:} For all $\stat_j$ in $\texttt{S}$, $\stat_j$ is neither in the $\irqset$ or $\dmaset$. i.e., $[(\stat_j \notin \irqset) \land (\stat_j \notin \dmaset)]  ~\lor~ (\stat_j \in \resetset)$.
					\end{compactenum}
					
				\end{compactitem}
			\end{compactitem}
			\vspace*{0.4em}
			\textbf{\ref{def:pfb_sec_def}.2 \pfb-Game:}
			The challenger plays the following game with \adv:
			\begin{compactenum}
				\item \adv is given full control over \dev software state, implying \adv can execute any (polynomially sized) sequence of arbitrary instructions $\{\inst^{\adv}_0, ..., \inst^{\adv}_n\}$, inducing the associated changes in \dev's sequence of execution states;
				\item \adv has oracle access to polynomially many calls to \verify. \adv also has access to the set of software executables, $\mathsf{SW} := \{S_1, ..., S_l\}$, and the set of all corresponding authorization ``tokens'', $\mathsf{T}:= \{\atoken_{1},...,\atoken_{l}\}$, ever produced by any prior \ctrl calls to \authorize up until time $t$. i.e., $\atoken_j \leftarrow \authorize(\fw_j, ...)$, for all $j$.
				\item Let $\mathsf{U} \subset \mathsf{T}$ be the set of all ``used'' authorization tokens up until time $t$, i.e., $\atoken_j \in \mathsf{U}$, if a call to $\privsense(\fw_j, ...)$ returned $\top$ up until time $t$; Let $\mathsf{P}$ be the set of ``pending'' (issued but not used) authorization tokens, i.e., $\mathsf{P} := \mathsf{T} \setminus \mathsf{U}$.
				\item At any arbitrary time $t$, \adv wins if it can perform an \textbf{unauthorized or tampered sensing execution}, i.e.:
				\begin{compactitem}[--]
					\item \adv triggers an $\privsense(\fw_{\adv},...)$ operation that returns $\top$, for $\forall \fw_{\adv} \notin \mathsf{SW}$, or
					\item \adv triggers $(\texttt{S}, \top) \leftarrow \privsense(\fw_j,...)$ such that $\atomic(\texttt{S}, \fw_j) \equiv \perp$, for some $\fw_j \in \mathsf{SW}$ and $\atoken_j \in \mathsf{U}$ . 
				\end{compactitem}	
			\end{compactenum}
			%
			\vspace*{0.4em}
			\textbf{\ref{def:pfb_sec_def}.3 \pfb-Security:}
			A scheme is considered \pfb-Secure iff, for all PPT adversaries \adv, there exists a negligible function $\negl[]$ such that: \vspace*{-0.5em}
			\begin{center}
				$Pr[\adv, \text{\pfb-Game}] \leq \negl[l]$
			\end{center}
		\end{definition}
	\end{mdframed}
	\vspace*{-1em}
\end{figure*}

Based on the specified machine model, we now proceed with the formal definition of \pfb.

\subsection{\pfb Syntax}\label{subsec:syntax}
A \pfb scheme involves two parties: \ctrl and \dev. \ctrl authorizes \dev to execute some software \fw which accesses \gpio.
It should be impossible for any software different from \fw to access \gpio data, or any function thereof (see  
Definition~\ref{def:pfb_sec_def}). \ctrl is trusted to only authorize functionally correct code. The goal of a \pfb scheme is to 
facilitate sensing-dependent execution while keeping all sensed data private from all other software.

Definition~\ref{def:pfb} specifies a syntax for \pfb scheme composed of three functionalities: \authorize, \verify, 
and \privsense. \authorize is invoked by \ctrl to produce an authorization token, \atoken, to be sent to \dev,  
enabling \fw to access \gpio. \verify is executed at \dev with \atoken as input, and
it checks whether \atoken is a valid authorization for the software on \dev. If and only if this check succeeds, 
\verify returns $\top$. Otherwise, it returns $\perp$. The verification success indicates one execution of 
\fw granted on \dev via \privsense. \privsense is considered successful 
(returns $\top$), if there is at least one MCU state produced by \privsense where a \gpio read 
occurs \underline{without} causing an MCU $reset$, i.e.,
	$(\stat \in \readset_{GPIO}) \land \neg(\stat \in \resetset)$.
Otherwise, \privsense returns $\perp$. That is, \privsense models execution of any software in the MCU and its 
return symbol indicates whether a \gpio read occurred during its execution. Therefore, invocation of \privsense 
on any input software that does not read from \gpio returns $\perp$. 
Figure~\ref{fig:pfb_interaction} illustrates a benign \pfb interaction between \ctrl and \dev.

\begin{figure}[t]
	\centering
	\scalebox{0.6}[0.5]{
		\fbox{
			\begin{tikzpicture}[node distance=1.5cm, >=stealth]
	\coordinate (BL)	at (0, 4.3);		\coordinate[left =6cm  of BL]	 (TL);
	\coordinate (Btvs)	at (0, 3.8);		\coordinate[left =6cm  of Btvs] (Ttvs);
	\coordinate (Btcs)	at (0, 2.8);		\coordinate[left =6cm  of Btcs] (Ttcs);
	\coordinate (Btce)	at (0, 1.3);		\coordinate[left =6cm  of Btce] (Ttce);
	\coordinate (Btvr)	at (0, 0.3);		\coordinate[left =6cm  of Btvr] (Ttvr);
	\coordinate (BR)	at (0, 0);		\coordinate[left =6cm  of BR]	 (TR);

	\node[above] at (BL) {\large \dev};
	\node[above] at (TL) {\large \ctrl};
	\coordinate (authorize) at ($(Btcs)$);
	\node (verify) [right = .1cm, align=center] at (authorize) {\small (2) Verify};
	\coordinate (execute) at ($(Btce)$);
	\node [right = .1cm, align=center] at (execute) {\small (3) Execute };

	\draw[line width = .2cm, color=gray!90]	(BL) -- (BR);
	\draw[line width = .2cm, color=gray!90] (TL) -- (TR);
	\coordinate (ReqStart) at ($(Ttvs)!0.1!(Btcs)$);
	\coordinate (ReqEnd) at ($(Ttvs)!0.9!(Btcs)$);
	\draw[thick, ->] (ReqStart) -- (ReqEnd) node [above=-.05cm, midway, sloped] {\small(1) Authorization};
	\coordinate (RepStart) at ($(Btce)!0.1!(Ttvr)$);
	\coordinate (RepEnd) at ($(Btce)!0.9!(Ttvr)$);
	\draw[thick, ->] (RepStart) -- (RepEnd) node [above=-.05cm, midway, sloped] {\small(4) Result (Optional)};
\end{tikzpicture}
		}
	}
	\vspace{-0.5em}
	\caption{\footnotesize \pfb interaction between \ctrl and \dev}\label{fig:pfb_interaction}
	\vspace{-0.8em}
\end{figure}

\subsection{\new{Assumptions \& Adversarial Model}}\label{sec:adv_model}
We consider an adversary, \adv, that controls the entire software state of \dev, including PMEM (flash) and DMEM (DRAM). 
It can attempt to modify any writable memory (including PMEM)
or read any memory, including peripheral regions, such as \gpio, unless explicitly protected by verified hardware.
It can launch code injection attacks to execute arbitrary instructions from PMEM or even DMEM (if the MCU architecture supports execution from DMEM).
It also has full control over any DMA controllers on \dev that can directly read/write to any part of the memory independently of the CPU.
It can induce interrupts to pause any software execution and leak information from its stack, or change its control-flow.
We consider Denial-of-Service (DoS) attacks, whereby \adv abuses \pfb functionality in order to render \dev unavailable, to be out-of-scope.
These are attacks on \dev availability and not on sensed data privacy. 

\new{
\textbf{Executable Correctness:} we stress that \pps aims to guarantee that \fw, as specified by 
\ctrl, is the only software that can access and process \gpio data. Similar to other trusted hardware architectures, 
\pfb does not check for lack of implementation bugs within \fw; thus it is not concerned with run-time (e.g., control-flow and data-only)
attacks. As a relatively powerful and trusted entity \ctrl can 
use various well-known vulnerability detection methods, e.g., fuzzing~\cite{fuzzer}, 
static analysis~\cite{costin2014large}, and even formal verification, to scrutinize \fw before authorizing it.
}

\new{
\textbf{Physical Attacks:} physical and hardware-focused attacks are considered out of scope. We assume 
that \adv cannot modify code in ROM, induce hardware faults, or retrieve \dev's secrets via side-channels that require 
\adv's physical presence. Protection against such attacks can be obtained via standard 
physical security techniques~\cite{ravi2004tamper}. This assumption is in line with related work on 
trusted hardware architectures for embedded systems~\cite{smart,vrasedp,trustlite,tytan}.
}

\subsection{\pfb Game-based Definition}\label{subsec:game_def}
Definition~\ref{def:pfb_sec_def} starts by introducing an auxiliary predicate \atomic.
It defines whether a particular sequence of execution states (produced by the execution of some software \fw)
adheres to all necessary execution properties for \emph{Atomic Sensing Operation Execution} discussed in Section~\ref{sec:overview}. 

In \atomic (in Definition~\ref{def:pfb_sec_def}.1), conditions 1-3 guarantee that a given \fw is executed as a whole and no external instruction is executed between its first and last instructions. Condition 4 assures that DMA is inactive during execution, hence protecting intermediate variables in DMEM against DMA tampering. Additionally, malicious interrupts could be leveraged to illegally change the control-flow of \fw during its execution. Therefore, condition 4 stipulates that both cases cause \atomic to return $\perp$.

\pfb-Game in Definition~\ref{def:pfb_sec_def}.2 models \adv's capabilities by allowing it to execute 
any sequence of (polynomially many) instructions. This models \adv's full control over software executed on the MCU, as well 
as its ability to use software to modify memory at will. It can also call \verify any 
(polynomial) number of times in an attempt to gain an advantage (e.g., learn something) from \verify executions.

To win the game, 
\adv must succeed in executing some software that does not cause an MCU reset, and either: (1) is unauthorized, 
yet reads from \gpio, or (2) is authorized, yet violates \atomic predicate conditions during its execution.

\section{\pps: Realizing \pfb} \label{sec:design}
\begin{figure}
   \centering
   \includegraphics[keepaspectratio=true, height=2.0in,width=0.9\columnwidth]{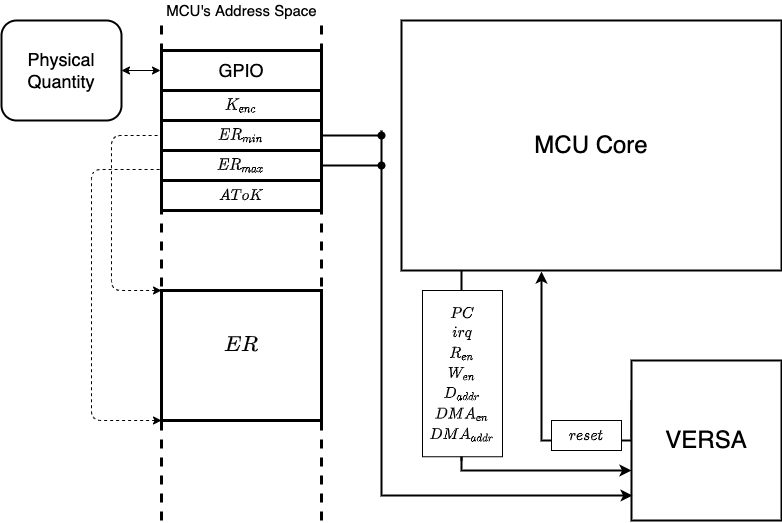}
   \vspace*{-0.7em}
   \caption{\footnotesize \pps Architecture} \label{fig:archpps}
   \vspace*{-1.7em}
\end{figure}
\pps runs in parallel with the MCU core and monitors a set of MCU signals: $PC$, $D_{addr}$, $R_{en}$, $W_{en}$, $DMA_{en}$, 
$DMA_{addr}$, and $irq$. It also monitors $ER_{min}$ and $ER_{max}$, the boundary memory addresses of $ER$ where 
\fw is stored; these are collectively referred to as ``\metadata''. \pps hardware module detects privacy violations in real-time, based on  
aforementioned signals and \metadata values, causing an immediate MCU reset.
Figure~\ref{fig:archpps} shows the \pps architecture. For quick reference, 
MCU signals and memory regions relevant to \pps are summarized in Table~\ref{tabl:notations}.
To facilitate specification of \pps properties, we introduce the following two macros:

\vspace{0.3em}
\noindent{\centering\footnotesize
	$Read\_Mem(i) \equiv (\ren \land \daddr = i) \lor (\dmaen \land \dmaaddr = i)$\\
	$Write\_Mem(i) \equiv (\wen \land \daddr = i) \lor (\dmaen \land \dmaaddr = i)$\\
}
\vspace{0.3em}

\noindent representing read/write from/to a particular memory address $i$ by either CPU or DMA. For reads/writes from/to 
some continuous memory region (composed of multiple addresses) $\mem = [\mem_{min},\mem_{max}]$, 
we instead say $D_{addr} \in \mem$ to denote that $D_{addr} = i ~\land (i \geq \mem_{min}) 
\land (i \leq \mem_{max})$. The same holds for notation $DMA_{addr} \in \mem$.

\begin{table}
\caption{\footnotesize Notation Summary} \label{tabl:notations}
\vspace{-1em}
\footnotesize
\linespread{0.8}
\renewcommand{\arraystretch}{1.1}
\begin{tabularx}{\columnwidth}{lX}
\toprule
Notation & Description    \\ 
\midrule
\pc		&	Current program counter value  \\
\ren 		&	1-bit signal that indicates if MCU is reading from memory  \\
\wen 	&	1-bit signal that indicates if MCU is writing to memory \\
\daddr 	&	Memory address of an MCU memory access   \\
\dmaen 	&	1-bit signal that indicates if DMA is active \\
\dmaaddr 	&	Memory address being accessed by DMA, when active     \\
\irq 		&	1-bit signal that indicates if an interrupt is happening   \\
\reset	&	Signal that reboots the MCU when set to logic `1'\\
\ER		&	A configurable memory region where the sensing operation \fw is stored, $\ER = [\ermin, \ermax]$\\
\metadata	&	Metadata memory region; contains \ermin and \ermax\\
\atoken	&	Fixed memory region from which \verify reads the authorization token when called \\
\gpio		&	Memory region that is mapped to GPIO port \\
\VR		&	Memory region storing \verify code which instantiates \vrased software and its hardware protection\\
\pass	&	A fixed address in \ROM, only be reachable (i.e., $PC=\pass$) by a successful \verify call (i.e., \verify 
			returns $\top$)\\
\ekr		&	(Optional) memory region for the encryption key \enckey necessary to encrypt the \fw output (relevant to  sensed data) \\
\bottomrule
\vspace*{-3em}
\end{tabularx}
\end{table}

\subsection{\pps: Construction} \label{subsec:construction}
\begin{figure*}
\begin{mdframed}
\label{def:versa}
\begin{construction}\label{cons:pps}
\footnotesize
\pps instantiates a \pfb = $[\authorize, \verify, \privsense]$ scheme as follows:
~\\~
-- \attkey is a symmetric key pre-shared between \ctrl and \vrased secure architecture in \dev;
\begin{compactenum}
~
\item{$\authorize^{\ctrl}(\fw)$:} \ctrl produces an authorization message $\texttt{M}:=(\fw, \chal, \atoken)$,
where \fw is a software, i.e., a sequence of instructions $\{\inst_1, ..., \inst_n\}$, that \ctrl wants to execute 
on \dev; \chal is a monotonically increasing challenge; and $\atoken$ is an authentication token computed as below.
\ctrl sends \texttt{M} to \dev. Upon receiving \texttt{M}, \dev is expected to parse \texttt{M}, 
find the memory region for \fw, and execute \verify (see below).
\begin{equation}
	\atoken := HMAC(KDF(\chal, \attkey), \fw)
\end{equation} 
\item{$\verify^{\dev}(ER, \atoken, \chal)$:} calls \vrased functionality~\cite{vrasedp} on memory 
region $ER:= [\ermin,\ermax]$ to securely compute: 
\begin{equation} \label{eq:verify_sigma}
\sigma := HMAC(KDF(\chal, \attkey), ER) 
\end{equation} 
\vspace*{-0.2em}
If $\sigma = \atoken$, output $\top$;
Otherwise, output $\perp$. \\
\item{$\privsense^{\dev}(ER)$:} 
starts execution of software in $ER$ by jumping to $ER_{min}$ (i.e., setting \pc = \ermin).
A benign call to \privsense with input \ER is expected to occur after one successful computation 
of \verify for the same $ER$ region and contents therein. Otherwise, \pps hardware support (see below) 
will cause the MCU to reset when \gpio is read. 
\privsense produces $\texttt{E}:=\{\stat_0, ..., \stat_m\}$, the set of states produced by executing $ER$, and outputs $\top$ or $\perp$ as follows:
\begin{equation}
\privsense(ER) = 
\begin{cases}
	(\texttt{E}, \top),& \text{ if } \exists \stat \in \texttt{E} \text{ such that } (\stat \in \readset_{\gpio}) \land (\stat \notin \resetset)\\
	(\texttt{E}, \perp),& \text{ otherwise}
\end{cases}
\end{equation}

\item \hwmon:
\underline{At all times}, \pps verified hardware enforces all following LTL properties :\\ \\
\linespread{0.5}
\footnotesize

{\bf A -- Read-Access Control to} \gpio{\bf:}
\begin{equation}\label{eq:read_gpio1}
\begin{split}
\text{\bf G}: \ \{
(Read\_Mem(\gpio) \land \neg(PC \in \ER)) \rightarrow \reset \}
\end{split}
\end{equation}
\begin{equation}\label{eq:read_gpio2}
\begin{split}
\text{\bf G}: \ \{ 
[(PC = \ermax) \lor \reset] \rightarrow (\neg Read\_Mem(\gpio) \lor \reset) ~\textbf{W}~ (PC = \pass) \}
\end{split}
\end{equation}
{\bf B -- Ephemeral Immutability of \ER and \metadata }
\begin{equation}\label{eq:auth_noWrite1}
\begin{split}
\text{\bf G}: \ \{(\pc=\pass) \land (Write\_Mem(\ER) \lor Write\_Mem(\metadata)) \rightarrow \reset \}
\end{split}
\end{equation}
\begin{equation}\label{eq:auth_noWrite2}
	\begin{split}
		\text{\bf G}: \{ ((Write\_Mem(\ER) \lor Write\_Mem(\metadata) \rightarrow (\neg Read\_Mem(\gpio) \lor \reset) 
		~\textbf{W}~ (PC = \pass))
		\}
	\end{split}
\end{equation}
\begin{equation}\label{eq:auth_noWrite3}
	\begin{split}
		\textnormal{[Optional] } \text{\bf G}: \{ ((Write\_Mem(\ER) \lor Write\_Mem(\metadata) \rightarrow 
		(\neg Read\_Mem(\ekr) \lor \reset) ~\textbf{W}~ (PC = \pass))
		\}
	\end{split}
\end{equation}
{\bf C -- Atomicity and Controlled Invocation of \ER:}
\begin{equation}\label{eq:ephe_continv1}
\begin{split}
\text{\bf G}: \ \{\neg \reset \land (PC \in \ER) \land  \neg \text{\bf X}(PC \in \ER) \rightarrow  (PC = \ermax) \lor \text{\bf X}(\reset) \}
\end{split}
\end{equation}
\begin{equation}\label{eq:ephe_continv2}
\begin{split}
 & \text{\bf G}: \ \{\neg \reset \land \neg (PC \in \ER) \land \text{\bf X}(PC \in \ER) \rightarrow \text{\bf X}(PC = \ermin) 
 \lor \text{\bf X}(\reset) \}
\end{split}
\end{equation}
\begin{equation}\label{eq:ephe_atom}
\begin{split}
 & \text{\bf G}: \ \{(PC \in \ER) \land (\irq \lor \dmaen) \rightarrow \reset \}
\end{split}
\end{equation}

{\bf [Optional] Read/Write-Access Control  to Encryption Key (\enckey) in \ekr:}
\begin{equation}\label{eq:read_enckey1}
\begin{split}
\text{\bf G}: \ \{
(Read\_Mem(\ekr) \land \neg(PC \in \ER)) \rightarrow \reset \}
\end{split}
\end{equation}
\begin{equation}\label{eq:read_enckey2}
\begin{split}
\text{\bf G}: \ \{
[(PC = \ermax) \lor \reset] 
\rightarrow (\neg Read\_Mem(\ekr) \lor \reset) ~\textbf{W}~ (PC=\pass) \}
\end{split}
\end{equation}
\begin{equation}\label{eq:write_enckey}
\begin{split}
\text{\bf G}: \ \{
[Write\_Mem(\ekr) \land \neg(PC \in \VR)] \rightarrow \reset \}
\end{split}
\end{equation}
\end{compactenum}

\textbf{\textit{Remark:}} [Optional] properties are needed only if support for encryption of outputs is desired.

\end{construction}
\end{mdframed}
\vspace{-0.8em}
\caption{\footnotesize \ppstext (\pps) Scheme}
\vspace{-0.8em}
\normalsize
\end{figure*}

Recall the key features of \pps from Section \ref{sec:overview}. To guarantee \emph{Mandatory Sensing Operation Authorization} 
and \emph{Atomic Sensing Operation Execution}, \pps constructs \pfb = $(\authorize, \verify, \privsense)$ algorithms as in 
Construction~\ref{cons:pps}.  We describe each algorithm below.

\begin{center}
\textbf{\authorize:}
\end{center}
\vspace{-.5em}
To authorize \fw, \ctrl picks a monotonically 
increasing \chal and generates $\atoken$ := $HMAC(KDF(\chal,\attkey),\fw)$
(this follows \vrased authentication algorithm -- see \pps~\verify specification below). \atoken is computed 
over \fw with a one-time key derived from \attkey and \chal, where \attkey is the master secret key shared between \ctrl and \dev.

\begin{center}
\textbf{\verify:}
\end{center}
\vspace{-.5em}
To securely verify that an executable \fw', installed in \ER, matches authorized \fw, \dev invokes \vrased\footnote{\dev and \ctrl act as \prover and \verifier in \vrased respectively.}
%
to compute
$\sigma $ := $HMAC(KDF(\chal,\attkey),\fw')$.
\verify outputs $\top$, iff $\sigma=\atoken$.
In this case, \pc reaches a fixed address, called \pass. Otherwise, \verify outputs $\perp$. 

In the rest of this section, we use ``authorized software" to refer to software located in \ER, for which 
$\verify(\ER, \atoken)$ outputs $\top$. Whereas, ``unauthorized software" 
refers to any software for which $\verify(\ER, \atoken)$ outputs $\perp$.

\begin{center}
\textbf{\privsense:}
\end{center}
\vspace{-.5em}
%
When \privsense(\ER) is invoked, \pc jumps to \ermin, and starts executing the code in \ER. 
It produces a set $\textit{\texttt{E}}$ of states by executing \ER, and outputs $\top$, if there is at least one state that 
reads \gpio without triggering an MCU reset. Otherwise, it outputs $\perp$.

\begin{center}
\textbf{\hwmon:}
\end{center}
\vspace{-.5em}

\pps \hwmon is verified to enforce LTL specifications (\ref{eq:read_gpio1})--(\ref{eq:write_enckey}) in Construction~\ref{cons:pps}.

{\bf A -- Read-Access Control to \gpio} is jointly specified by LTLs (\ref{eq:read_gpio1}) and (\ref{eq:read_gpio2}). LTL (\ref{eq:read_gpio1}) 
states that \gpio can only be read during execution of \ER ($\pc \in \ER$), requiring an MCU reset otherwise. LTL (\ref{eq:read_gpio2}) 
forbids all \gpio reads (even those within $ER$ execution) before successful computation of \verify on $ER$ binary using a valid 
\atoken. Successful \verify computation is captured by condition $PC=\pass$. A new successful computation of $\verify(ER,\atoken)$ 
is necessary whenever $ER$ execution completes ($PC=ER_{max}$) or after reset/boot. Hence, each legitimate \atoken 
can be used to authorize $ER$ execution once. 

{\bf B -- Ephemeral Immutability of \ER and \metadata} is specified by LTLs (\ref{eq:auth_noWrite1})-(\ref{eq:auth_noWrite3}).
From the time when $ER$ binary is authorized until it starts executing, 
no modifications to \ER or \metadata are allowed.
LTL (\ref{eq:auth_noWrite1}) specifies that no such modification is allowed at the moment when verification succeeds ($\pc = \pass$);
LTL (\ref{eq:auth_noWrite2}) requires $ER$ to be re-authorized from scratch if $ER$ or \metadata are ever modified. Whenever these modifications are detected $(Write\_Mem(\ER) \lor Write\_Mem(\metadata)$) further reads to \gpio are immediately blocked ($\neg Read\_Mem(\gpio) \lor \reset$) until subsequent re-authorization of $ER$ is completed ($ ... ~\textbf{W}~ (PC = \pass)$). 
LTL (\ref{eq:auth_noWrite3}) specifies the same requirement in order to read \pps-provided encryption key (\enckey) which is stored in memory region \ekr. This property is only required when support for encryption of outputs is desired.

{\bf C -- Atomicity \& Controlled Invocation of \ER} are enforced by LTLs (\ref{eq:ephe_continv1}),  (\ref{eq:ephe_continv2}), and~(\ref{eq:ephe_atom}). They specify that \ER execution must start at
\ermin and end at \ermax. Specifically, they use the relation between \emph{current} and \emph{next} \pc values.
The only legal \pc transition from currently outside of \ER to next inside \ER is via $PC=\ermin$. Similarly, the only legal \pc transition from currently inside \ER to next outside \ER
is via $PC=\ermax$. All other cases trigger an MCU reset. In addition, LTL (\ref{eq:ephe_atom}) requires an MCU reset whenever interrupts or 
DMA activity is detected during \ER execution. This is done by simply checking \irq and \dmaen signals.

\new{
We note that \privsense relies on the \hwmon to reset the MCU when violations to \ER atomic execution are detected. Upon reset all data is erased from memory.
However, when execution of \fw completes successfully \pps does not trigger resets. In this case, \fw is responsible for erasing its own stack before completion (reaching of $ER$ its last instruction).
We discuss how this self-clean-up routine can be implemented as a part of \fw behavior in Appendix \ref{apdx:clean_up}.
}

\subsection{Encryption \& Integrity of \ER Output} \label{subsec:enc_key}
As mentioned in Section~\ref{sec:overview}, after reading and processing \gpio inputs, \fw might need to encrypt and send the result to \ctrl. \pps supports encryption of this output, regardless of the underlying encryption scheme. For that purpose, 
\verify implementation derives a fresh one-time encryption key (\enckey) from \attkey and \chal. 
To assure confidentiality of \enckey, the following properties are required for the memory region (\ekr) reserved to store \enckey:
\begin{compactenum}
	\item \ekr is writable only by \verify (i.e., $\pc \in \VR$); and
	\item \ekr is readable only by \ER after authorization.
\end{compactenum}
LTLs (\ref{eq:read_enckey1})-(\ref{eq:write_enckey}) and (\ref{eq:auth_noWrite3}) specify the confidentiality requirements of \enckey.
In sum, these properties establish the same read access-control policy for \ekr and \gpio regions.
Therefore, only authorized \fw is able to retrieve \enckey.

\section{Verified Implementation \& Security Analysis} \label{sec:security_impl}
\subsection{Sub-module Implementation \& Verification} \label{subsec:submodule_imp_vrf}

\pps sub-modules are represented as FSMs and individually verified to hold for LTL properties from 
Construction~\ref{cons:pps}. They are implemented in Verilog HDL as Mealy machines, i.e., 
their output is determined by both their current state and current inputs.
Each FSM has a single output: a local \reset. \pps global output \reset is given by the disjunction (logic $OR$) 
of all local \reset-s. For simplicity, instead of explicitly representing the output \reset value for each state, we use 
the following convention:
\begin{compactenum}
	\item \reset is $1$ whenever an FSM transitions to $RESET$ state;
	\item \reset remains $1$ while on $RESET$ state;
	\item \reset is $0$ otherwise.
\end{compactenum}
Note that all FSMs remain in $RESET$ state until $\pc = 0$ which indicates that 
the MCU reset routine finished.

\begin{figure}
	\centering
	\includegraphics[width=1\columnwidth]{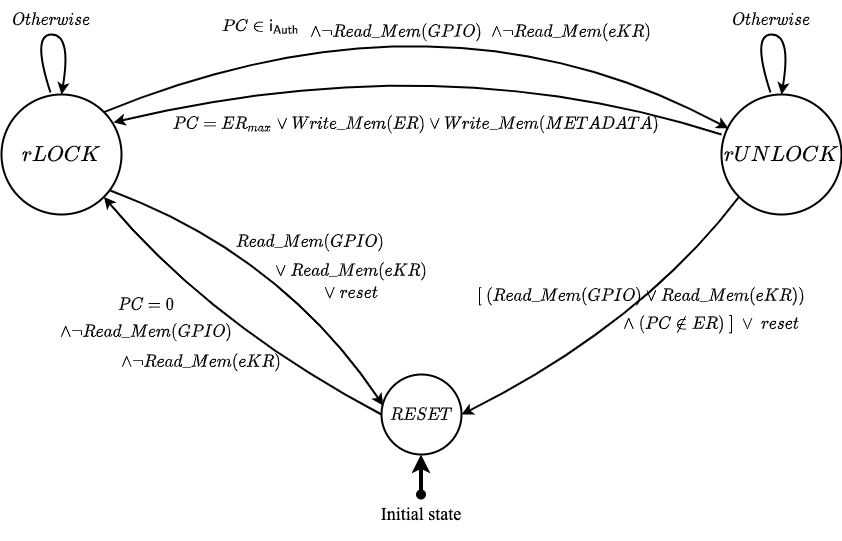}
	\vspace{-0.8em}
	\caption{\footnotesize Verified FSM for \gpio and \ekr Read-Access Control (LTL (\ref{eq:read_gpio1})-(\ref{eq:auth_noWrite3}) 
	\& LTL (\ref{eq:read_enckey1})-(\ref{eq:read_enckey2}))}  
	\label{fig:readprotection_gpio}
	\vspace{-0.8em}
\end{figure}

Fig.~\ref{fig:readprotection_gpio} illustrates the \pps sub-module that implements read-access control to \gpio and \ekr (when applicable). It guarantees that such reads are only possible when they emanate from execution of authorized software \fw contained in \ER. It also assures that no modifications to \ER or \metadata occur between authorization of \fw and its subsequent execution. The Verilog implementation of this FSM is formally verified to adhere to LTLs (\ref{eq:read_gpio1})-(\ref{eq:auth_noWrite3}) and (\ref{eq:read_enckey1})-(\ref{eq:read_enckey2}). It has 3 states: 
(1) $rLOCK$, when reads to \gpio (and possibly \ekr) are disallowed; 
(2) $rUNLOCK$, when such reads are allowed to \ER; and (3) $RESET$. 
The initial state (after reset or boot) is $RESET$, and it switches to $rLOCK$ state when $\pc = 0$. 
It switches to $rUNLOCK$ when $\pc=\pass$ (with no reads to \gpio and \ekr), indicating that 
\verify was successful. Note that $rUNLOCK$ transitions to $RESET$ when reads are attempted from outside 
\ER, thus preventing reads by any unauthorized software. Once \pc reaches \ermax, indicating that \ER execution has finished, the FSM transitions back to $rLOCK$. 
Also, any attempted modifications to \metadata or \ER in $rUNLOCK$ state bring the FSM back to $rLOCK$. 
Note that $rUNLOCK$ is only reachable after authorization of $ER$, i.e., $\pc=\pass$.

\begin{figure}
	\centering
	\includegraphics[width=1\columnwidth]{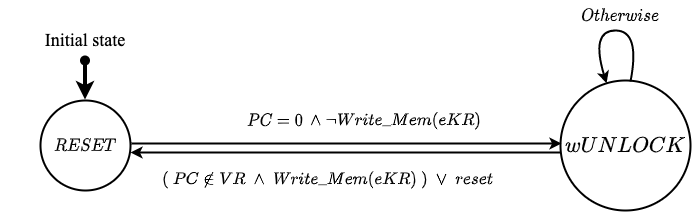}
	\vspace{-0.7em}
	\caption{\footnotesize Verified FSM for \ekr Write-Access Control (LTL (\ref{eq:write_enckey}))}
	\vspace{-0.8em}
	\label{fig:writeprotection_ekr}
\end{figure}
\begin{figure}
	\centering
	\includegraphics[width=1\columnwidth]{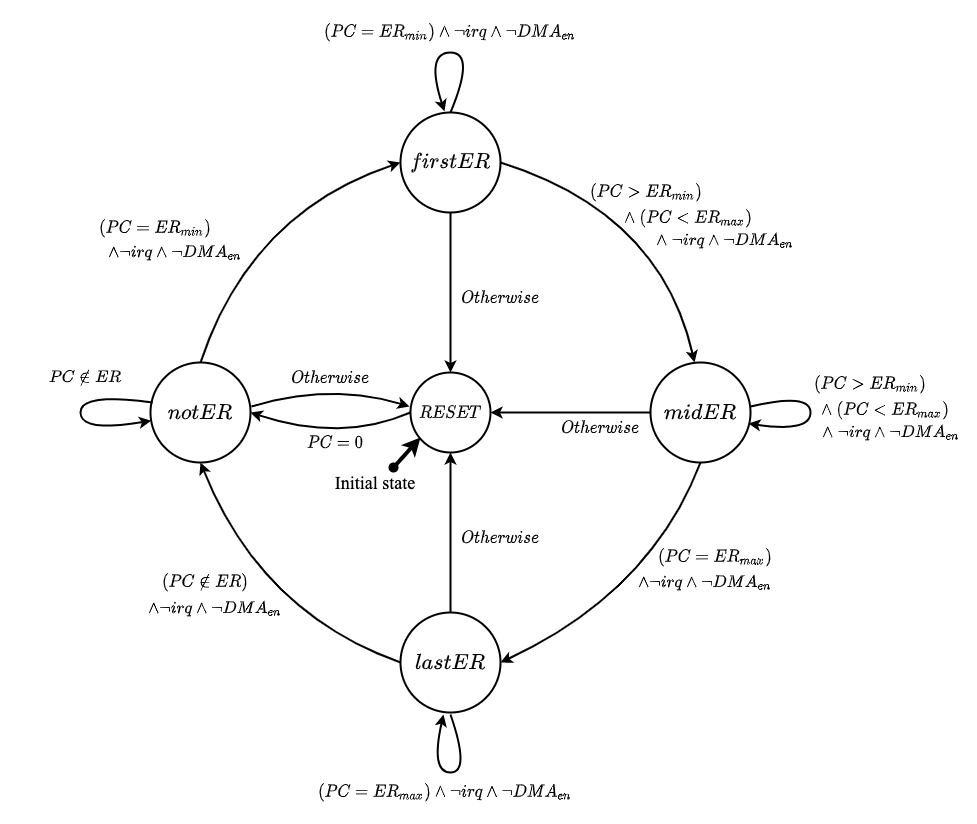}
	\vspace{-1em}
	\caption{\footnotesize \ER Atomicity and Controlled Invocation FSM (LTL (\ref{eq:ephe_continv1})-(\ref{eq:ephe_atom}))}
	\vspace{-0.8em}
	\label{fig:atomicex_er}
\end{figure}

\begin{figure*} 
\begin{mdframed}
\footnotesize
\begin{definition} \label{def:e2e_1_atom} \underline{Atomic Sensing Operation Execution:} \scriptsize
\begin{align*}
\begin{split}
	&\quad \textbf{G} \{
~
			(\pc \in \ER) \rightarrow [(\pc \in \ER) \land \neg \irq \land \neg \dmaen] ~\textbf{W}~ [(\pc = \ermax) \lor \reset] 
~			\} \\
	&\land \quad
	\textbf{G} \{
	~
			\neg\reset \land \neg(\pc \in \ER) \land \textbf{X}(\pc \in \ER) \rightarrow \textbf{X}(\pc = \ermin) \lor \textbf{X}(\reset)
	~
			\}
\end{split}
\end{align*}
\end{definition}
\begin{definition}\label{def:e2e_2_sensing} \underline{Mandatory Sensing Operation Authorization:}
\scriptsize 
\begin{align*}
\begin{split}
	& \textbf{G} \{~(Read\_Mem(\gpio) \land \neg \reset) \rightarrow (\pc \in \ER)~\} ~\land \\
	& 
	\Big\{  
	(\pc = \pass) \land 
	\big\{(\pc = \pass) \rightarrow [\neg Write\_Mem(\ER) \land \neg Write\_Mem(\metadata) \land (Write\_Mem(\ekr) \rightarrow (\pc \in \VR))]~\textbf{U}~(\pc = \ermin) \big\} \\
	& \Big\} 
	 ~\textbf{B}~ \big\{Read\_Mem(\gpio) \land \neg \reset\big\}
\end{split}
\end{align*}
\end{definition}	
\end{mdframed}
\vspace{-2em}
\end{figure*}

The FSM in Figure~\ref{fig:writeprotection_ekr} enforces LTL (\ref{eq:write_enckey}) to protect \ekr from 
external writes. It has two states:  
(1) $wUNLOCK$, when writes to \ekr are allowed; and (2) $RESET$. 
At boot/after reset ($PC=0$), this FSM transitions from $RESET$ to $wUNLOCK$. It transitions back to 
$RESET$ state whenever writes to \ekr are attempted, unless these writes come from \verify execution ($\pc \in \VR$). 

Figure~\ref{fig:atomicex_er} shows the FSM verified to enforce \ER atomicity and controlled invocation: LTLs (\ref{eq:ephe_continv1})-(\ref{eq:ephe_atom}). It has five states; $notER$ and $midER$ correspond to 
\pc being outside and within \ER (not including \ermin and \ermax), respectively.
$firstER$ and $lastER$ are states in which \pc points to \ermin and \ermax, respectively.
The only path from $notER$ to $midER$ is via $firstER$. Likewise, the only path from $midER$ to $notER$ is via
$lastER$. The FSM transitions to $RESET$ whenever \pc transitions do not follow 
aforementioned paths. It also transitions to $RESET$ (from any state other than $notER$) if $\irq$ or $\dmaen$ signals are set. 
\subsection{Sub-module Composition and \pps End-To-End Security}\label{subsec:composition_e2e}
To demonstrate security of \pps according to Definition~\ref{def:pfb_sec_def}, our strategy is two-pronged:
\begin{compactenum}
	\item[\bf A)] We show that LTL properties from Construction~\ref{cons:pps} are sufficient to imply that \gpio  (and \ekr)
	is only readable by \fw and 
	any \privsense operation that returns $\top$ (i.e., performs sensing) is executed atomically. 
	The former is formally specified in Definition~\ref{def:e2e_2_sensing}, and the latter in Definition~\ref{def:e2e_1_atom}. For this part, we write an LTL computer 
	proof using SPOT LTL proof assistant~\cite{spot}.
	\item[\bf B)] We use a cryptographic reduction to show that, as long as item \textbf{A} holds, \vrased security 
	can be reduced to \pps security according to Definition~\ref{def:pfb_sec_def}. 
\end{compactenum}
The intuition for this strategy is that, to win \pfb-game in Definition~\ref{def:pfb_sec_def}, \adv must either break the atomicity of 
\privsense (which is in direct conflict with Definition~\ref{def:e2e_1_atom}) or execute \privsense with unauthorized 
software and read \gpio without causing an MCU \reset. Definition~\ref{def:e2e_2_sensing} guarantees that the 
latter is not possible without a prior successful call to \verify. On the other hand, \verify is implemented using \vrased 
verified architecture, which guarantees the unforgeability of \atoken. Hence, breaking \pps requires either violating 
\pps verified guarantees or breaking \vrased verified guarantees, which should be infeasible to any PPT \adv.

Appendix~\ref{apdx:proofs} includes proofs for Theorems~\ref{thm:atom_ER},~\ref{thm:readGPIO}, and~\ref{thm:reduction}, 
in accordance to this proof strategy. The rest of this section focuses on \pps end-to-end implementation goals captured by 
LTLs in Definitions~\ref{def:e2e_1_atom} and~\ref{def:e2e_2_sensing} as well as their relation to \pps 
high-level features discussed in Section~\ref{sec:overview}.

\begin{mdframed}
\footnotesize
\begin{theorem}\label{thm:atom_ER}
	Definition \ref{def:hw_model} $\land$ LTL \ref{eq:ephe_continv1}, \ref{eq:ephe_continv2}, \ref{eq:ephe_atom} $\rightarrow$ Definition \ref{def:e2e_1_atom}
\end{theorem}
\vspace{-0.5em}
\begin{theorem}\label{thm:readGPIO}
	Definition \ref{def:hw_model} $\land$ LTL \ref{eq:read_gpio1}, \ref{eq:read_gpio2}, \ref{eq:auth_noWrite1}, \ref{eq:auth_noWrite2}, \ref{eq:ephe_continv2}, \ref{eq:write_enckey}  $\rightarrow$ Definition \ref{def:e2e_2_sensing}
\end{theorem}
\vspace{-1em}
\begin{theorem}\label{thm:reduction}
	\pps is secure according to the \pfb-game in Definition~\ref{def:pfb_sec_def}, as long as \vrased is a secure \RA architecture according to \vrased security game from \cite{vrasedp}.
\end{theorem}
\end{mdframed}

\vspace{1mm} \noindent\emph{\textbf{[Definition~\ref{def:e2e_1_atom}]}} states that it {\bf globally} (always) holds 
that \ER is atomically executed with controlled invocation.
That is, whenever an instruction in \ER executes ($\pc \in \ER$),  it keeps executing instructions within \ER ($\pc \in \ER$), 
with no interrupts and no DMA enabled, {\bf until} \pc reaches the last instruction in \ermax or an MCU reset occurs. Also, if an instruction in \ER starts to execute, it always begins with the first instruction in $\ermin$. 
This formally specifies the \textit{Atomic Sensing Operation Execution} feature discussed in Section~\ref{sec:overview}.

\vspace{1mm} \noindent\emph{\textbf{[Definition~\ref{def:e2e_2_sensing}]}} {\bf globally} requires that whenever \gpio is successfully read 
(i.e., without a \reset), this read must come from the CPU while $ER$ is being executed. 
In addition, {\bf before} this read operation, the
following must have happened at least once:
\begin{compactitem}
	\item[(1)] \verify succeeded (i.e., $\pc = \pass$);
	\item[(2)] From the time when $\pc = \pass$ {\bf until} \ER starts executing (i.e., $\pc = \ermin$), no modification to 
	\ER and \metadata occurred; and
	\item[(3)] If there was any write to \ekr from the time when $\pc = \pass$, {\bf until} $\pc = \ermin$, it 
	must have been from \verify, i.e., while $\pc \in \VR$. 
\end{compactitem}
This formally specifies the intended behavior of the  \emph{Mandatory Sensing Operation Authorization} feature, 
discussed in Section~\ref{sec:overview}.

\section{Evaluation \& Discussion} \label{sec:evaluation}
In this section, we discuss \pps implementation details and evaluation. \pps source code and verification/proofs are publicly available at \cite{pfb-repo}.
Evaluation of verification costs and discussion of \pps limitations are deferred to Appendix \ref{apdx:eval}.
\subsection{Toolchain \& Prototype Details}
\pps is built atop OpenMSP430 \cite{openmsp430}: an open source implementation of TI-MSP430 \cite{TI-MSP430}. 
We use Xilinx Vivado to synthesize an 
RTL description of \hwmon and deploy it on Diligent Basys3 prototyping board for Artix7 FPGA. For the software part 
(mostly to implement \verify), \pps extends \vrased software (which computes $HMAC$ over \dev memory) to include 
a comparison with the received \atoken (See Section~\ref{subsec:runtime_overhead} for extension details). 
%
%
We use the 
NuSMV model checker to formally verify that \hwmon implementation adheres to LTL specifications (\ref{eq:read_gpio1})-(\ref{eq:write_enckey}). See Appendix~\ref{apdx:eval} for details on the verification setup and costs.


\subsection{Hardware Overhead}
Table~\ref{tab:overhead_results} reports on \pps hardware overhead, as compared to unmodified OpenMSP430 and \vrased. 
Similar to other schemes~\cite{vrasedp,apex,sancus,smart}, we consider hardware overhead in terms of additional 
Look-Up Tables (LUTs) and registers. Extra hardware in terms of LUTs gives an estimate of additional chip cost and size 
required for combinatorial logic, while extra hardware in terms of registers gives an estimate of memory overhead required by 
sequential logic in \pps FSMs. Compared to \vrased, \pps requires 10\% additional LUTs and 2\% additional registers. 
In actual numbers, it adds 255 LUTs and 50 registers to the underlying MCU as shown in Table~\ref{tab:overhead_results}. 

\begin{table}[!htp]
\footnotesize
\centering
\vspace{-1em}
\caption{\footnotesize \small Hardware Overhead and Verification cost}
\vspace{-0.8em}
	\resizebox{\columnwidth}{!}{ 
		\begin{tabular}{|l|cc|c|cccc|} \hline\cline{1-8}
			\multirow{2}{*}{Architecture} & \multicolumn{2}{c|}{Hardware} & Reserved &\multicolumn{4}{c|}{Verification}      \\
			& LUTs            & Regs     & RAM~(bytes)      & LoC & \#(LTLs) &Time~(s) & RAM~(MB) \\ \hline
			OpenMSP430                & 1854           & 692     & 0      & -    &   -     & -        & -           \\
			\vrased                   & 1891           & 724      & 2332     & 481    &  10    & 0.4      &  13.6       \\
			\pps + \vrased                 & 2109           & 742       & 2336    & 1118     &  21   & 13956.4      &  1059.1       \\
			\hline\cline{1-8}
		\end{tabular}%
	}
	\vspace{-0.3cm}
	\label{tab:overhead_results}
\end{table}

\begin{figure}
	\centering 
	\subfigure[Additional HW overhead (\%) in Number of Look-Up Tables]
	{\includegraphics[width=0.49\columnwidth]{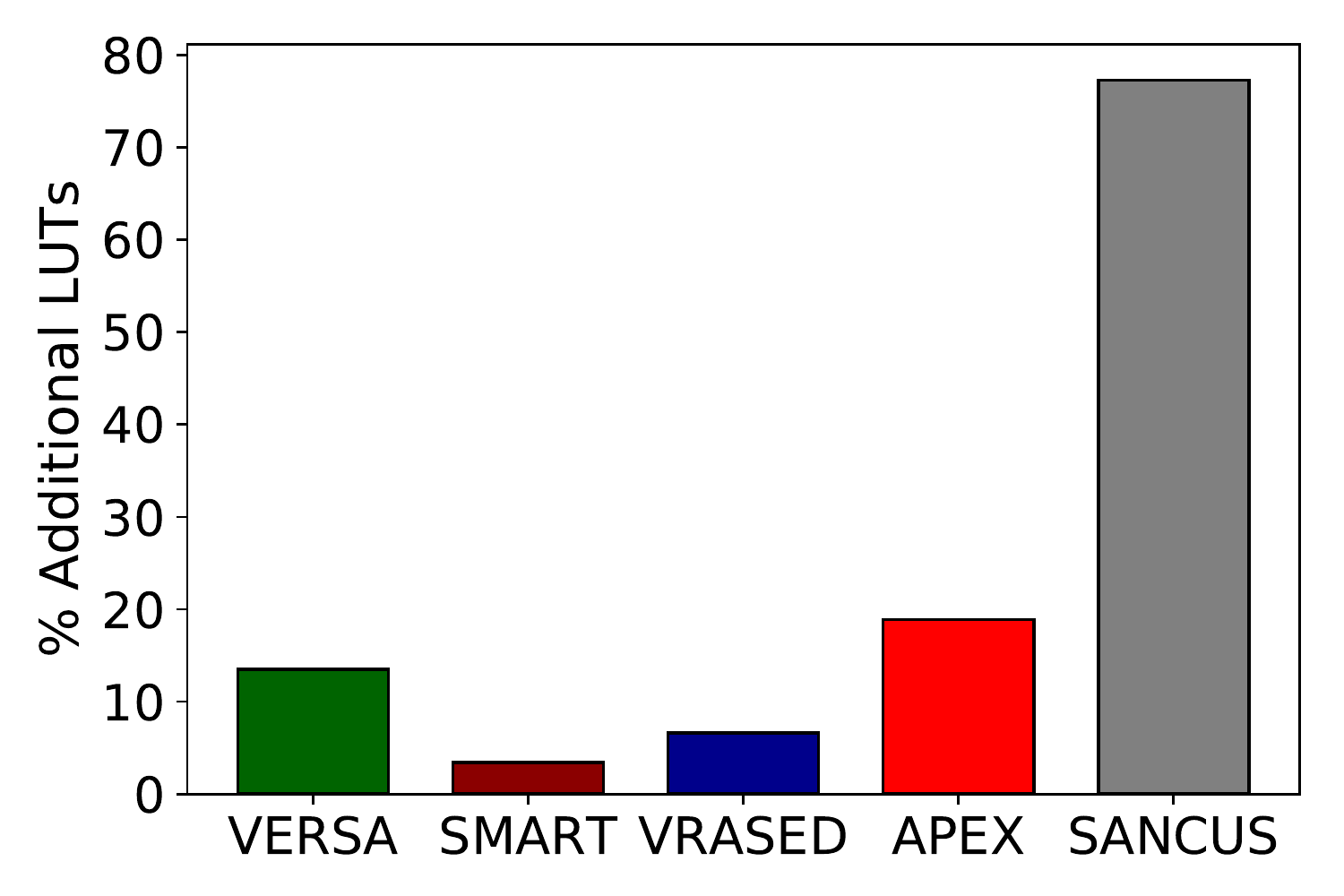}}
	\subfigure[Additional HW overhead (\%) in Number of Registers]
	{\includegraphics[width=0.49\columnwidth]{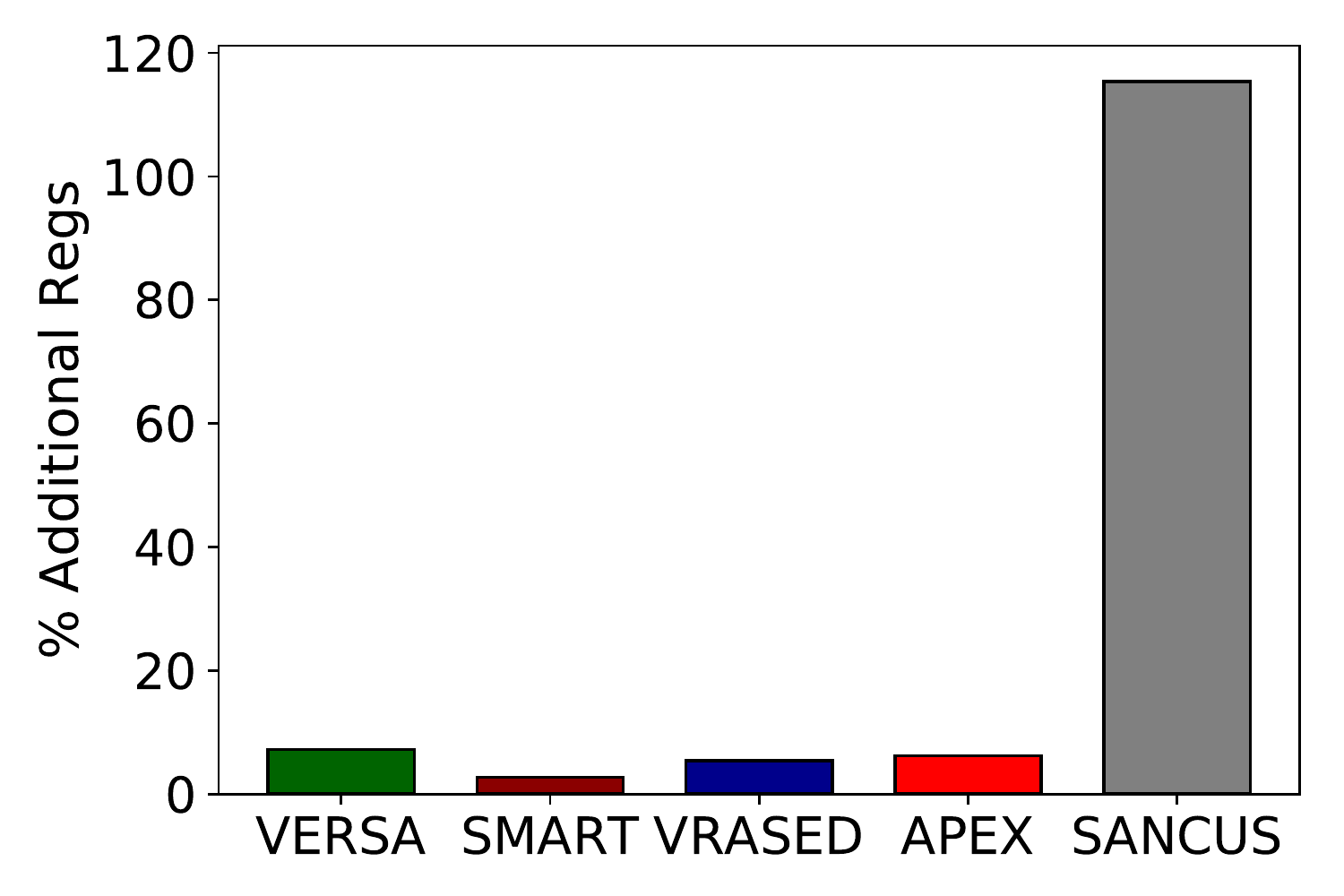}}
	\vspace{-1.4em}
	\caption{\footnotesize Hardware overhead comparisons with other low-end security architectures.}
	\label{fig:comparison}
	\vspace{-1.5em}
\end{figure}

\subsection{Runtime Overhead} \label{subsec:runtime_overhead}

\begin{figure}
	\centering
	\includegraphics[width=0.6\columnwidth]{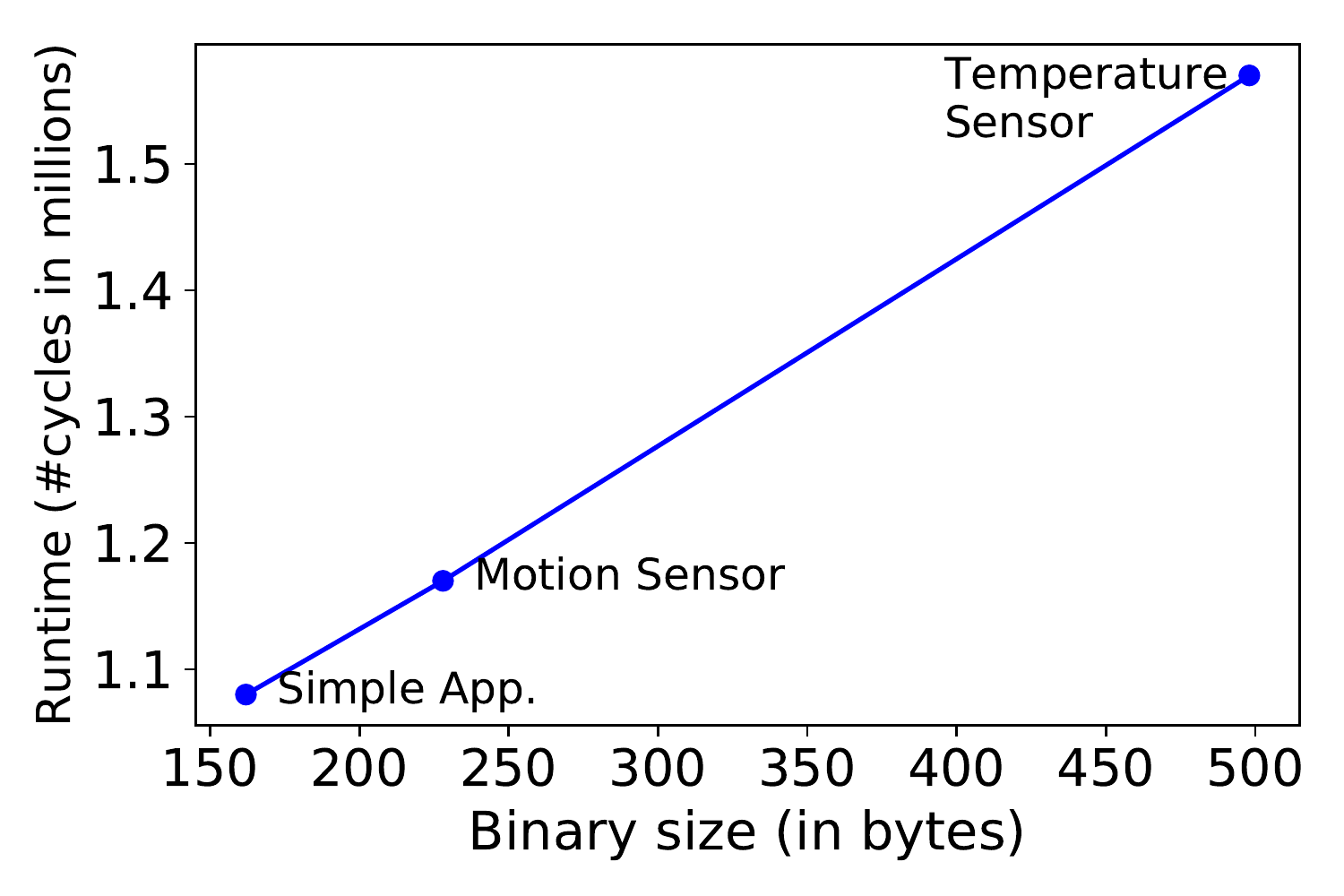}
	\vspace{-1.4em}
	\caption{\footnotesize Runtime overhead of \pps due to \verify} \label{fig:runtime_overhead}
	\vspace{-1.5em}
\end{figure}

\pps requires any software piece seeking to access \gpio (and \enckey) to be verified. Consequently, runtime overhead is due to \verify computation which instantiates \vrased. This runtime includes: {(\bf1)} time to compute 
$\sigma$ from equation \ref{eq:verify_sigma}; {(\bf2)} time to check if $\sigma = \atoken$; and {(\bf3)} time to write \enckey to 
\ekr, when applicable. Naturally the runtime overhead is dominated by the computation in {\bf (1)} which is proportional to the size of $ER$. 

We measure \verify cost on three sample applications: (1) Simple Application, which reads 32-bytes of 
\gpio input and encrypts it using One-Time-Pad (OTP) with \enckey; (2) Motion Sensor -- available at~\cite{code1} -- 
which continuously reads \gpio input to detect movements and actuates a light source when movement is detected; and 
(3) Temperature Sensor -- adapted code from~\cite{code2} to support encryption of its outputs -- 
which reads ambient temperature via \gpio and encrypts this reading using OTP. 
We prototype using OTP for encryption for the sake of simplicity noting that \pps does not mandate a particular encryption scheme. 
All of these sample applications also include a self-clean-up code executed immediately before reaching their exit point to erase their stack traces once their execution is over.

Figure \ref{fig:runtime_overhead} shows \verify runtimes on these applications. Assuming a clock frequency of 10MHz (a common frequency for low-end MCUs), \verify runtime ranges from $100-200$ milliseconds for these applications. The overhead is linear on the binary size.


%
\subsection{Comparison with Other Low-End Architectures:} 
To the best of our knowledge, \pps is the first architecture related to \pfb. 
However, to provide a point of reference in terms of performance and overhead, we compare \pps with other low-end 
trusted hardware architectures, such as SMART~\cite{smart}, VRASED~\cite{vrasedp}, APEX~\cite{apex}, and 
SANCUS~\cite{sancus}. All these architectures provide \RA-related services to attest integrity of software 
on \dev either statically or at runtime. Since \pfb also checks software integrity before granting access to \gpio, we consider these architectures to be related to \pps.

Figure \ref{fig:comparison} compares \pps hardware overhead with the aforementioned architectures in terms of additional LUTs and registers. Percentages are relative to the plain MSP430 core total cost. 

\pps builds on top of \vrased. As such, it is naturally more expensive than hybrid \RA architectures such as SMART and \vrased.
Similar to \pps, APEX also monitors execution properties and also builds on top of \RA (in APEX case, with the goal of producing proofs of remote software execution). Therefore, \pps and APEX exhibit similar overheads. SANCUS presents a higher cost because it implements \RA and isolation features in hardware. 

\section{Related Work} \label{sec:rw}
There is a considerable body of work (overviewed in Section~\ref{sec:intro}) on IoT/CPS privacy. However,
to the best of our knowledge, this paper is the first effort specifically targeting \pfb, i.e., sensor data privacy on potentially compromised MCUs. 
Nonetheless, prior work has proposed trusted hardware/software 
co-designs -- such as \pps~-- offering other security services. We overview them in this section.

Trusted components, commonly referred to as Roots of Trust (RoTs), are categorized as software-based,
hardware-based, or hybrid (i.e., based on hardware/software co-designs).
Their usual purpose is to verify software integrity on a given device.
Software-based RoTs~\cite{KeJa03,seshadri2004swatt,SLS+05,SLP08,GGR09,LMP11,gligor} usually do not rely on any hardware modifications. 
However, they are insecure against compromises to the entire software state of a device (e.g., in cases where \adv can physically re-program \dev). 
In addition, their inability to securely store cryptographic secrets imposes reliance on strong assumptions about precise timing and 
constant communication delays to enable device authentication. These assumptions can be unrealistic in the IoT ecosystem.
Nonetheless, software-based RoTs are the only viable choice for legacy devices that have no security-relevant hardware support.
Hardware-based methods~\cite{PFM+04,TPM,KKWHAB2012,SCHELLEKENS200813,flicker,MQY10,sancus} rely on security
provided by dedicated hardware components (e.g., TPM~\cite{TPM} or ARM TrustZone~\cite{trustzone}). However,
the cost of such hardware is normally prohibitive for low-end MCUs. Hybrid RoTs~\cite{smart,apex,vrasedp,tytan,trustlite}
aim to achieve security equivalent to hardware-based mechanisms, yet with lower hardware costs.
They leverage minimal hardware support while relying on software to reduce additional hardware complexity.

\new{Other architectures, such as SANCTUM \cite{sanctum} and Notary \cite{notary}, provide strong 
memory isolation and peripheral isolation guarantees, respectively.
These guarantees are achieved via hardware support or external hardware agents. 
They are also hybrid architectures where trusted hardware works in tandem with trusted software.
However, we note that such schemes are designed for application computers that support MMUs and are therefore unsuitable for simple MCUs.}

In terms of functionality, such embedded RoTs focus on integrity. Upon receiving a request from an external trusted {\it Verifier},
they can generate unforgeable proofs for the state of the MCU or that certain actions were performed by the MCU.
Security services implemented by them include: (1) memory integrity verification, i.e., remote attestation
~\cite{smart,sancus,vrasedp,simple,tytan,trustlite}; 
(2) verification of runtime properties, including control-flow and data-flow attestation
~\cite{apex,litehax,cflat,lofat,atrium,oat,tinycfa,dialed,geden2019hardware};
and (3) proofs of remote software updates, memory erasure, and system-wide resets~\cite{pure,verify_and_revive,asokan2018assured}.
As briefly mentioned in Section~\ref{sec:intro}, 
due to their reactive nature, 
they can be used to \emph{detect} whether \dev has been compromised \emph{after the fact}, 
but \emph{cannot prevent} the compromised entity from exfiltrating private sensor data. \pps, on the other hand,
 enforces mandatory authorization before any sensor data access and thus prevents leakage even when a compromise has already happened.

Formalization and formal verification of RoTs for MCUs have gained attention due
to the benefits discussed in Sections~\ref{sec:intro} and~\ref{sec:BG}. VRASED~\cite{vrasedp}
implemented a verified hybrid \RA scheme. APEX~\cite{apex} built atop
VRASED to implement and formally verify an architecture for proofs of remote execution of
attested software. PURE~\cite{pure} implemented provably secure services for software update, memory erasure, and system-wide reset. Another recent result~\cite{busi2020provably}
formalized and proved the security of a hardware-assisted mechanism to prevent leakage of secrets through timing side-channels due to MCU interrupts. Inline with
the aforementioned work, \pps also formalizes its assumptions along with its goals and implements the
first formally verified design assuring \pfb.

\vspace*{-0.2em}

\section{Conclusions} \label{sec:conclusion}
We formulated the notion of \pfbtext (\pfb) and proposed \pps: a formally verified architecture realizing \pfb.
\pps ensures that only duly authorized software can access sensed data even if the entire software state of the sensor
is compromised. To attain this, \pps enhances the underlying MCU with a small hardware monitor, which is shown sufficient 
to achieve \pfb. The experimental evaluation of \pps publicly available prototype~\cite{pfb-repo} demonstrates 
its affordability on a typical low-end IoT MCU: TI MSP430.

\section*{Acknowledgments}
The authors sincerely thank S\&P'22 reviewers. This work was supported by funding 
from NSF Awards SATC-1956393 and CICI-1840197, as well as a subcontract from Peraton Labs. 
Part of this work was conducted while the first author was at the University of California, Irvine.

\bibliographystyle{./IEEEtran}
\bibliography{./references}

\begin{appendix}
\subsection{\new{Clean-up after Program Termination}}\label{apdx:clean_up}

\new{
	While \pps guarantees the confidentiality of sensing operations, it requires the authorized executable \fw to erase its own stack/heap before its termination.
	This ensures that unauthorized software can not extract 
	and leak sensitive information from \fw execution and allocated data. 
	Erasure in this case can be achieved via a single call to libc's \texttt{memset} function with start address matching the base of \fw stack and size equal to the maximum size reached by \fw execution.
}

\begin{figure}
	\centering
	\includegraphics[width=1\columnwidth]{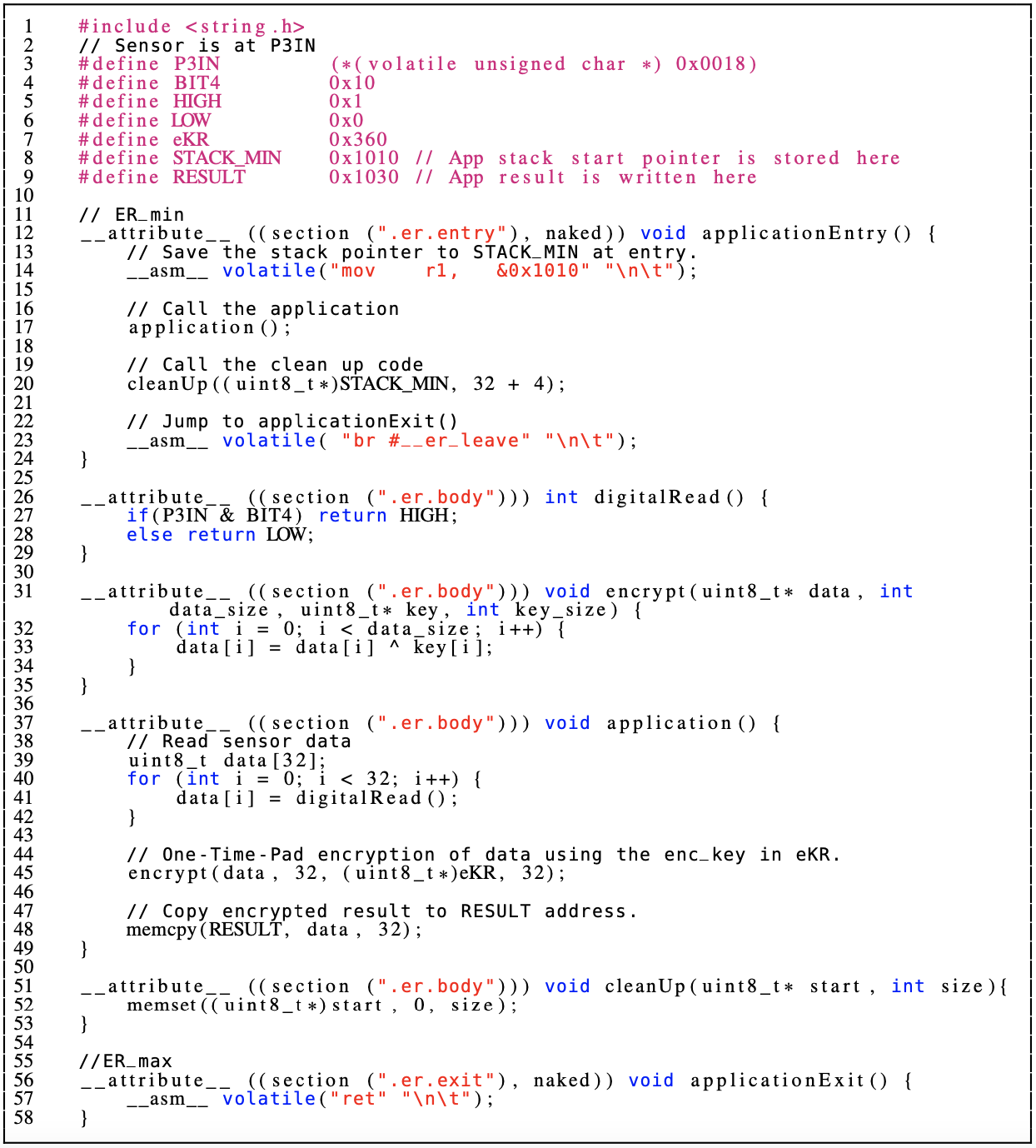}
	\vspace{-1.8em}
	\caption{\footnotesize \new{Sample sensing operation that reads GPIO input, encrypts it, and cleans up its stack after execution.}}
	\vspace{-0.8em}
	\label{fig:app}
\end{figure}

\new{
	The maximum stack size can be determined manually by counting the allocated local variables in small and simple \fw implementations. 
	To automatically determine this size in more complex \fw implementations, all functions called within \fw must update the highest point reached by their respective stacks. 
	Figure \ref{fig:app} shows a sample application that reads 32 bytes of sensor data, encrypts this data using \pps one-time key \enckey,
	and cleans-up the stack thereafter.
	Line 12 is \fw entry point (\ermin).
	\fw first saves the stack pointer to \texttt{STACK\_MIN} address. Then, the \texttt{application} function is called, in line 17. \texttt{application} implements \fw intended behavior. After the \texttt{application} is done, the clean up code (lines 51-53) is called with \texttt{STACK\_MIN} as the start pointer and size of 32 + 4 bytes 
	(32 bytes for \texttt{data} variable (in line 39) and 4 bytes for stack metadata).
}

\subsection{\new{Data Erasure on Reset/Boot}}\label{apdx:erasure}
\new{
	Violations to \pps properties trigger an MCU reset.
	A reset
	immediately stops execution and prepares the 
	MCU core to reboot by clearing all
	registers
	and pointing 
	the program counter (\pc) 
	to the first instruction of PMEM.
	However, some MCUs may not guarantee erasure of DMEM as a part of this process.
	Therefore, traces of data allocated by \fw (including sensor data) could persist across resets.
}

\new{
	In MCUs that do not offer DMEM erasure on reset, a software-base Data Erasure (\DE)
	can be implemented and invoked it as soon as the MCU starts, i.e., as a part of the bootloader code. 
	In particular, \DE can be implemented using \texttt{memset} (similar to lines 51-53) with constant arguments matching the entirety of the MCU's DMEM.
	\DE should be immutable (e.g., stored in ROM) which is often the case for bootloader binaries.
	Upon reset, \pc must always point to the first instruction of \DE. The normal MCU start-up proceeds normally after \DE execution is completed.
}

\subsection{\pps Composition Proof}\label{apdx:proofs}

In this section, we show that \pps is a secure \pfb architecture according to Definition~\ref{def:pfb_sec_def}, as long as \textbf{A)} the sub-properties in Construction~\ref{cons:pps} hold (Theorem~\ref{thm:atom_ER}, \ref{thm:readGPIO}) and \textbf{B)} \vrased is a secure remote attestation (\RA) architecture according to the \vrased security definition in \cite{vrasedp} (Theorem~\ref{thm:reduction}). 
Informally, part \textbf{A)} shows that if the machine model and all LTLs in Construction~\ref{cons:pps} hold, then the end-to-end goals for secure \pfb architecture are met, while this does not include the goal of prevention of forging authorization tokens. Part \textbf{B)} handles the latter using a cryptographic reduction, i.e., it shows that an adversary able to forge the authorization token (with more than negligible probability) can also break \vrased according to the \RA-game, which is a contradiction assuming the security of \vrased. Therefore, Theorems~\ref{thm:atom_ER}-\ref{thm:reduction} prove that \pps is a secure \pfb architecture as long as \vrased is a secure \RA architecture.

For part \textbf{A)}, computer-checked LTL proofs are performed using SPOT LTL proof assistant~\cite{spot}. These proofs are available at~\cite{pfb-repo}. We present the intuition behind them below.\\

\noindent \textit{Proof of Theorem~\ref{thm:atom_ER} (Intuition).}
LTL (\ref{eq:ephe_continv2}) states the legal entry instruction requirement, while LTL (\ref{eq:ephe_continv1}) states the legal exit instruction requirement in \atomic. Also, since LTL (\ref{eq:ephe_continv1}) states that \ermax is the only possible exit of the \ER without a reset, it implies self-contained execution of \ER. Lastly, LTL (\ref{eq:ephe_atom}) enforces MCU reset if any interrupt or DMA occurs, which naturally prevents interrupts and DMA actions, as required by \atomic. These imply the LTL in Definition~\ref{def:e2e_1_atom} which stipulates that execution of \ER must start with \ermin and stays within \ER with no interrupts nor DMA actions, until \pc reaches \ermax (causing a reset otherwise). \qed \\

\noindent \textit{Proof of Theorem~\ref{thm:readGPIO} (Intuition).}
Definition~\ref{def:e2e_2_sensing} {\bf (i)} requires at least one successful verification of \ER  before  \gpio can be read successfully (without triggering a reset); and {\bf (ii)} disallows modifications to \ER, \metadata, and \enckey (other than by \VR) in between \ER verification subsequent \ER execution.
LTLs (\ref{eq:read_gpio1}) and (\ref{eq:read_gpio2}) state that \pc must be within \ER to read \gpio and disallow \gpio reads by default (including when MCU reset occurs) and after the execution of \ER is over ($\pc = \ermax$). Also, LTL (\ref{eq:read_gpio2}) requires (re-)authorization ($\pc = \pass$) of \ER after the execution of \ER is over ($\pc = \ermax$). 
LTL (\ref{eq:auth_noWrite2}) disallows \gpio reads until the (re-)verification whenever \ER or \metadata are written. LTL (\ref{eq:auth_noWrite1}) disallows changes to \ER and \metadata at the exact time when verification succeeds. LTL (\ref{eq:ephe_continv2}) guarantees that the execution of \ER starts with \ermin and LTL (\ref{eq:write_enckey}) guarantees that only the \VR code can modify the value in \ekr. Thus, these are sufficient imply Definition~\ref{def:e2e_2_sensing}. \qed \\

For part \textbf{B)}, we construct a reduction from the security game of \vrased in \cite{vrasedp} to the security game of \pps according to the Definition~\ref{def:pfb_sec_def}. i.e., The ability to break the \pfb-game of \pps allows to break the \RA-game of \vrased, and therefore, as long as \vrased is a secure \RA architecture according to the \RA-game, \pps is secure according to the \pfb-game.\\

\noindent \textit{Proof of Theorem~\ref{thm:reduction}.}  
Denote by $\adv_{\pfb}$, an adversary who can win the security game in Definition~\ref{def:pfb_sec_def} against \pps with more than negligible probability. We show that if such $\adv_{\pfb}$ exists, then it can be used to construct $\adv_{\RA}$ that wins the \RA-game with more than negligible probability. 

Recall that, to win the \pfb-game, $\adv_{\pfb}$ must trigger $\top$ as a result of \privsense, which means it reads the sensed data without MCU reset. From the \pfb-game step 4 in Definition~\ref{def:pfb_sec_def}, it can be done in either of the following two ways:

\noindent \texttt{Case1.} $\adv_{\pfb}$ executes a new, unauthorized software $\fw_{\adv}$ which causes $\privsense(\fw_{\adv}) \rightarrow \top$; or \\
\texttt{Case2.} $\adv_{\pfb}$ breaks the atomic execution of an authorized, but not yet executed software, $\fw_j$, so that it causes $\privsense(\fw_j) \rightarrow (E, \top)$ such that $\atomic(E, \fw_j) \equiv \perp$.

Recall that for the instruction set $I_j$ of $\fw_j$ and a set $E_j$ of execution \operations, to have $\atomic(I_j, E_j) \equiv \perp$, at least one of four requirements in Definition~\ref{def:pfb_sec_def}.1 must be false.
Note that the atomic sensing operation execution goal in Definition~\ref{def:e2e_1_atom} rules out the probability of \texttt{Case2}. Specifically, LTL~(\ref{eq:ephe_continv2}) enforces 1), while 2) and 3) are guaranteed by LTL~(\ref{eq:ephe_continv1}). Lastly, 4) is covered by LTL~(\ref{eq:ephe_atom}).

For \texttt{Case1}, i.e., to trigger $\top$ by running \privsense, $\adv_{\pfb}$ needs to read \gpio without causing an MCU reset. Recall 
that the \textit{Mandatory Sensing Operation Authorization} in Definition~\ref{def:e2e_2_sensing} requires \verify (with input executable in \ER)
to succeed at least once before reading \gpio. According to \pps construction, $\adv_{\pfb}$ 
causes $\verify(\ER, \atoken^*, \chal^*)$ to output $\top$, where \ER contains $\fw_{\adv}$ which is an unauthorized software, 
$\atoken^*$ is a valid issued (but never used) token, and $\chal^*$ is its corresponding challenge.
Since \verify function is implemented using \vrased to compute HMAC of \chal and \ER, such $\adv_{\pfb}$ can be directly used as $\adv_{\RA}$ to win the \RA-game of \vrased. 
Thus, assuming secure \RA architecture \vrased, this is a contradiction, which implies the security of \pps according to the \pfb-game.
\qed

\subsection{Extended Evaluation \& Discussion}\label{apdx:eval}
\vspace{1mm}
\noindent{\bf Verification Costs:}
Formal verification costs are reported in Table~\ref{tab:overhead_results}. 
We use a Ubuntu 18.04 desktop machine running at 3.4GHz with 32GB of RAM for formal verification.
Our verification pipeline converts Verilog HDL to SMV specification language and then verifies it against the LTL properties listed in Construction \ref{cons:pps} using the NuSMV model checker (per Section~\ref{sec:BG}).
\pps verification requires checking 11 extra invariants -- LTLs (\ref{eq:read_gpio1}) to (\ref{eq:write_enckey}) -- in addition to \vrased LTL invariants. 
It also incurs higher run-time and memory usage than \vrased verification. This is due to two additional 16-bit hardware signals (\ermin, \ermax) which increase the space of possible input combinations and thus
the complexity of model checking process. However, verification is still 
manageable in a commodity desktop -- it takes around 5 minutes and consumes 340MB of memory.

\vspace{3mm}
\noindent{\bf \pps Limitations:}

%

\subsubsection{Shared Libraries} 
to verify \fw, \ctrl must ensure that \fw spans one contiguous memory region (\ER) on \dev. 
If any code dependencies exist outside of \ER, \pps will reset the MCU according to LTL (\ref{eq:ephe_atom}). To preclude this
situation, \fw must be made self-contained by statically linking all of its dependencies within $ER$. 

\subsubsection{Atomic Execution \& Interrupts} 
per Definition~\ref{def:pfb_sec_def}, \pps forbids interrupts during execution of \privsense. This can be problematic,
especially on a \dev with strict real-time constraints. In this case, \dev must be reset in order to allow servicing the interrupt after DMEM erasure. This can cause a delay that could be harmful to real-time settings.
Trade-offs between privacy and real-time constraints should be carefully considered when using \pps. One possibility to remedy 
this issue is to allow interrupts as long as all interrupt handlers are: (1) themselves immutable and uninterruptible from 
the start of \privsense until its end; and (2) included in \ER memory range and are thus checked by \verify. 

\subsubsection{Possible Side-channel Attacks}
MSP430 and similar MCU-s allow configuring some GPIO ports to trigger interrupts. If one of such ports is used for triggering 
an interrupt, \adv could possibly look at the state of \dev and learn information about \gpio data. For example, suppose that 
a button press mapped to a GPIO port triggers execution of a program that sends some fixed number of packets over the 
network. Then, \adv can learn that the GPIO port was activated by observing network traffic. To prevent such attacks, privacy-sensitive quantities should always be physically 
connected to GPIO ports that are not interrupt sources 
(these are usually the vast majority of available GPIO ports). Other popular timing attacks related to cache side-channels and 
speculative execution, are not applicable to this class of devices, as these features are not present in low-end MCUs.

\subsubsection{\new{Flash Wear-Out}}
\new{
\pps implements \verify using \vrased. As discussed in Section~\ref{subsec:BG-vrased}, the  authentication protocol suggested by \vrased requires persistent storage of the highest value of a monotonically increasing challenge/counter in flash. 
We note that flash memory has a limited number of write cycles (typically at least 10,000 cycles~\cite{msp_memory_specs,atmel_specs}). Hence, a large number of successive counter updates may wear-out flash causing malfunction.
In \vrased authentication, the persistent counter stored in flash is only updated following successful authentication of \ctrl. Therefore, only legitimate requests from \ctrl cause these flash writes, limiting the capability of an attacker to exploit this issue.
Nonetheless, if the number of expected legitimate calls to \verify is high, one must select the persistent storage type or (alternatively) use different flash blocks once a given flash block storing the counter reaches its write cycles' limit. For a more comprehensive discussion of this matter, see~\cite{microvault}. 
}

\vspace{3mm}
\noindent{\bf \pps Alternative Use-Case:}

\noindent \pps can be viewed as a general technique to control access to memory regions based on software authorization tokens. 
We apply this framework to \gpio in low-end MCUs.
Other use-cases are possible.
For example, a \pps-like architecture could be used to mark a secure storage region and grant access 
only to explicitly authorized software. This could be useful if \dev runs multiple (mutually distrusted) applications and data must be securely shared between subsets thereof.

\end{appendix}

\end{document}